\documentclass[12pt]{article}

\usepackage{latexsym}
\usepackage{amssymb,amsfonts,amsmath}
\usepackage{graphicx} 
\usepackage{indentfirst}
\usepackage{bbm}
\usepackage{amssymb}
\usepackage{verbatim}
\usepackage{amsmath, amsthm,amssymb}
\usepackage{mathrsfs}
\usepackage{hyperref}
\usepackage{amsfonts}
\usepackage{dsfont}
\usepackage{cite}
\usepackage{xcolor}
\usepackage{enumerate}

\newcommand\Tstrut{\rule{0pt}{2.6ex}}         
\newcommand\Bstrut{\rule[-0.9ex]{0pt}{0pt}}   

\parskip=\medskipamount

\newcommand {\cD}{{\cal D}}
\newcommand {\cE}{{\cal E}}

\newcommand {\cL}{{\cal L}}

\newcommand {\cN}{{\cal N}}

\newcommand {\cQ}{{\cal Q}}
\newcommand {\cR}{{\cal R}}

\newcommand {\cV}{{\cal V}}

\def\a{\alpha}

\def\b{\beta}
\def\c{\chi}
\def\d{\delta}
\def\e{\epsilon}
\def\f{\phi}
\def\g{\gamma}
\def\G{\Gamma}

\def\j{\psi}
\def\k{\kappa}
\def\l{\lambda}
\def\m{\mu}

\def\o{\omega}

\def\q{\theta}
\def\r{\rho}
\def\s{\sigma}
\def\t{\tau}

\def\x{\xi}
\def\z{\zeta}
\def\D{\Delta}
\def\F{\Phi}
\def\J{\Psi}

\def\O{\Omega}
\def\P{\Pi}

\def\U{\Upsilon}
\def\X{\Xi}

\def\rd{{\rm d}}
\def\ri{{\rm i}}
\def\re{{\rm e}}

\newcommand{\ad}{{\dot{\alpha}}}                           
\newcommand{\bd}{{\dot{\beta}}}                            
\newcommand{\ve}{\varepsilon}                            
\newcommand{\cDB}{{\bar\cD}}                            

\newcommand{\pa}{\partial}                           
\newcommand{\hf}{\frac12}

\newcommand{\vf}{\varphi}

\newcommand{\be}{\begin{equation}}
\newcommand{\ee}{\end{equation}}
\newcommand{\bea}{\begin{eqnarray}}
\newcommand{\eea}{\end{eqnarray}}
\newcommand{\non}{\nonumber}
\def\double #1{#1{\hbox{\kern-2pt $#1$}}}

\newcommand{\gd}{{\dot\g}}

\newcommand{\qb}{{\bar{\theta}}}

\newif\ifdtup

\newcommand{\bsubeq}{\begin{subequations}}
\newcommand{\esubeq}{\end{subequations}}

\newcommand{\mub}{{{\bar{\mu}}}}

\numberwithin{equation}{section}

\newcommand{\sSO}{\mathsf{SO}}

\newcommand{\sOSp}{\mathsf{OSp}}

\begin{document}

\begin{titlepage}
\begin{flushright}
 January, 2021 \\
 Revised version: April, 2021
\end{flushright}
\vspace{5mm}

\begin{center}
{\LARGE \bf AdS superprojectors}
\\ 
\end{center}

\begin{center}

{\bf
E. I. Buchbinder, D. Hutchings, S. M. Kuzenko and  M. Ponds} \\
\vspace{5mm}

\footnotesize{
{\it Department of Physics M013, The University of Western Australia\\
35 Stirling Highway, Perth W.A. 6009, Australia}}  
~\\

\vspace{2mm}
~\\
\texttt{Email: evgeny.buchbinder@uwa.edu.au, daniel.hutchings@research.uwa.edu.au,
sergei.kuzenko@uwa.edu.au, michael.ponds@research.uwa.edu.au}
\vspace{2mm}

\end{center}

\begin{abstract}
\baselineskip=14pt
Within the framework of ${\cal N}=1$ anti-de Sitter (AdS) supersymmetry in four dimensions,
we derive superspin projection operators (or superprojectors). For a  tensor superfield 
$\mathfrak{V}_{\alpha(m)\dot \alpha (n)}  := \mathfrak{V}_{(\alpha_1...\alpha_m)
(\dot \alpha_1...\dot \alpha_n)}$ on AdS superspace, with $m$ and $n$ non-negative integers, the corresponding superprojector turns 
$\mathfrak{V}_{\alpha(m)\dot \alpha(n)} $ into a multiplet
with the properties of a conserved conformal supercurrent. 
It is demonstrated that the poles of such superprojectors correspond to (partially) massless multiplets, and the associated gauge transformations are derived. 
We give a systematic discussion of how to realise the unitary and the partially massless representations of the ${\cal N}=1$ AdS${}_4$ superalgebra $\mathfrak{osp} (1|4)$ in terms of on-shell superfields. As an example, we present an off-shell model for the massive gravitino multiplet in AdS$_4$. We also prove that the gauge-invariant actions for superconformal higher-spin multiplets factorise into products of minimal second-order differential operators. 
\end{abstract}
\vspace{5mm}

\vfill

\vfill
\end{titlepage}

\newpage
\renewcommand{\thefootnote}{\arabic{footnote}}
\setcounter{footnote}{0}

\tableofcontents{}
\vspace{1cm}
\bigskip\hrule

\allowdisplaybreaks


\section{Introduction} \label{section1}

Superprojectors \cite{SalamS,Sokatchev,SG,Sokatchev81,RS,GGRS,BHHK} are superspace
projection operators which single out irreducible representations of supersymmetry.  
Various applications of such operators have appeared in the literature, 
including the following constructions: 
 (i) superfield equations of motion \cite{OS1,OS2}; (ii)  gauge-invariant actions
in four dimensions \cite{GS,GKP}; and (iii)  off-shell  $\cal N$-extended superconformal actions in three dimensions \cite{BHHK}.\footnote{Ref. \cite{BHHK} was a natural extension of the work \cite{BKLP}, where the spin projection operators  in three dimensions were constructed and used to obtain simple expressions for the higher-spin Cotton tensors. } Of special interest are those superprojectors 
which single out the highest superspin of tensor superfields 
$ \mathfrak{V}_{\a(m)\ad(n)} (x,\q,\bar \q)  := 
\mathfrak{V}_{\a_1...\a_m \ad_1...\ad_n}(x,\q,\bar \q)=
\mathfrak{V}_{(\a_1...\a_m)(\ad_1...\ad_n)}(x,\q,\bar \q)$, since they 
may be viewed as supersymmetric extensions of the 
Behrends-Fronsdal spin projection operators \cite{BF,Fronsdal}. 
We recall that a tensor field $\vf_{\a(m) \ad(n)} (x)$ of Lorentz type  $(m/2,n/2) $ is mapped by the corresponding Behrends-Fronsdal projector\footnote{Refs. \cite{BF,Fronsdal} made use of the four-vector notation in conjunction with the four-component spinor formalism, which resulted in rather complicated expressions for the spin projection operators.
However, switching to the two-component spinor formalism leads to 
remarkably simple and compact expressions for these projectors \cite{SG,GGRS}.} 
to a transverse field $\vf^{\rm T}_{\a(m) \ad(n)} (x)$ 
such that 
\bea
\pa^{\b \bd} \vf^{\rm T}_{\b \a(m-1) \bd \ad(n-1)} =0~.
\eea
This constraint is the characteristic feature of a conserved 
current $j_{\a(m) \ad(n)} $.

The four-dimensional results of \cite{SalamS,Sokatchev,SG,Sokatchev81,RS,GGRS}
correspond to the Poincar\'e supersymmetry.  Not much is known about the structure of 
superprojectors corresponding to the AdS$_4$ supersymmetry $\sOSp(1|4)$. More precisely, forty years ago 
Ivanov and Sorin \cite{IS} introduced the so-called transverse linear and longitudinal linear superfields on $\cN=1$ AdS superspace AdS${}^{4|4}$
and presented the projectors which single out such supermultiplets. 
 We recall that a complex tensor superfield $ \G_{\a(m)\ad(n)} $ on  AdS${}^{4|4}$
 is said to be {\it transverse linear} if it satisfies the constraint
\bea
\cDB^\bd \G_{\a(m)\bd \ad(n-1)} &=& 0 \quad \Longleftrightarrow \quad (\cDB^2-2(n+2)\mu)\G_{\a(m)\ad(n)} = 0~, \qquad n>0~,
\label{1.1}
\eea
where  $\m\neq 0$ is a constant parameter which determines the curvature of AdS$^{4|4}$,
see section \ref{superspace} below. 
A complex tensor superfield $G_{\a(m)\ad(n)}$ is said to be {\it longitudinal linear} if it satisfies the constraint
\bea 
\cDB_{(\ad_1}G_{\a(m)\ad_2\dots\ad_{n+1})}&=&0 \quad \Longleftrightarrow \quad (\cDB^2+2n\mu)G_{\a(m)\ad(n)}=0~.
\label{1.2}
\eea
For $n=0$ the first constraint in \eqref{1.1} is not defined, while the second condition 
\bea
(\cDB^2-4\mu)\G_{\a(m)} = 0
\label{1.3}
\eea
defines a linear superfield. For $n=0$ the constraint \eqref{1.2} defines a chiral superfield. 
The constraints \eqref{1.1}, \eqref{1.2} and \eqref{1.3} are the only differential constraints in AdS${}^{4|4}$ which define off-shell supermultiplets with unconstrained component fields \cite{IS}.
Certain transverse linear and longitudinal linear supermultiplets 
play a fundamental role 
in the massless supersymmetric higher-spin gauge theories in AdS${}^{4|4}$ \cite{KS94} and their ancestors in Minkowski superspace \cite{KPS,KS}, see also \cite{Sibiryakov} for a review.

Ref.  \cite{IS} described the projectors to the spaces of superfields which are constrained by  \eqref{1.1}, \eqref{1.2} and \eqref{1.3}, see eq. \eqref{TLLL} below. However, these are not 
supersymmetric extensions of the Behrends-Fronsdal spin projection operators
\cite{BF,Fronsdal}, since all independent component fields of the resulting superfield are unconstrained. Supersymmetric spin
projection operators are required to turn an unconstrained superfield
on AdS$^{4|4}$ into one with the properties of a conserved current supermultiplet. 
It is pertinent to describe the structure of conformal supercurrents in AdS${}^{4|4}$ following the more general  analysis of conserved current supermultiplets in a supergravity background
\cite{KR}. 

Let $m $ and $ n$ be positive integers.
A tensor superfield $J_{\a(m) \ad(n)}$ on AdS$^{4|4}$
is called a conformal supercurrent of Lorentz type  $(m/2,n/2)$ 
 if it obeys the two constraints
\begin{subequations}\label{supercurrent} 
\bea
\cD^{\b} J_{\b\a (m-1)\ad(n)} &=& 0 \quad \Longleftrightarrow \quad \big( \cD^{2} - 2 ( m + 2 ) \bar{\m} \big) J_{\a(m) \ad(n)} = 0 ~, 
\label{supercurrent-a} 
\\
\bar \cD^{\bd} J_{\a(m) \bd \ad(n-1) } &=& 0 \quad \Longleftrightarrow \quad \big( \cDB^{2} - 2 ( n + 2 ) \m \big) J_{\a(m) \ad(n)} = 0 ~.
\label{supercurrent-b} 
\eea
\end{subequations}
If $m=n$,  it is consistent to 
restrict  $ J_{\a(n) \ad(n)} $ to be real, $\bar J_{\a(n) \ad(n)} = J_{\a(n) \ad(n)}$.
The $m=n=1$ case corresponds to the ordinary 
conformal supercurrent \cite{FZ}. The case $m=n>1$ was first described in Minkowski superspace in \cite{HST81} (see also \cite{KMT})
and extended to AdS$^{4|4}$ in \cite{BHK}. The  case $m=n+1>1$ was first described in Minkowski superspace in \cite{KMT}
and extended to AdS$^{4|4}$ in \cite{BHK}.

If $m >n =0$, the constraints \eqref{supercurrent} should be replaced with 
\begin{subequations}\label{supercurrent2} 
\bea
\cD^{\b} J_{\b\a (m-1)} &=& 0 \quad \Longleftrightarrow \quad 
\big( \cD^{2} - 2 ( m + 2 ) \bar{\m} \big) J_{\a(m)} = 0 ~, 
\label{supercurrent2-a} 
\\
(\bar \cD^2 -4\m)  J_{\a(m)  } &=& 0~.
\label{supercurrent2-b} 
\eea
\end{subequations}
The $m=1$  case was first considered in \cite{KT}, where it was shown that the spinor supercurrent $J_\a$ naturally originates from the reduction of the conformal $\cN=2$ supercurrent \cite{Sohnius79}
to $\cN=1$ superspace. 

Finally, for $m=n=0$  
the constraints \eqref{supercurrent2} should be replaced with 
\begin{subequations}\label{supercurrent3}
\bea 
(\cD^2 - 4\bar \m) J &=&0~, \\
(\bar \cD^2 -4\m)  J &=& 0~.
\eea
\end{subequations}
These constraints describe a  flavour current supermultiplet \cite{FWZ} in AdS${}_4$. 
Irreducible supermultiplets of the types \eqref{supercurrent}, \eqref{supercurrent2}  and 
\eqref{supercurrent3} have been used in \cite{BKS} for the covariant quantisation of 
the massless supersymmetric higher-spin gauge theories in AdS${}_4$ \cite{KS94}.

The Behrends-Fronsdal projectors have  been generalised to the case of AdS${}_4$ only recently in \cite{KP20}. One of the important outcomes of \cite{KP20} was a new understanding of the so-called partially massless fields in AdS$_4$. Such fields
in diverse dimensions were  studied earlier in \cite{DW2, DeserN1, DeserN2, DeserW1, DeserW3, DeserW4, Higuchi1, Higuchi2, Higuchi3, Zinoviev,Metsaev2,SV,Metsaev,BG}.
Specifically, it was shown in \cite{KP20} that the partially massless fields 
are associated with the poles of the spin projection operators in AdS$_4$. 
 In the present paper we provide an $\cN=1$ supersymmetric extension of the AdS$_4$ spin projection operators and describe various applications of the resulting superprojectors.
In particular,  we will demonstrate that the partially massless $\cN=1$ supermultiplets in AdS$_4$, which have recently been analysed in the component settings in 
\cite{G-SHR,BKSZ}, are naturally associated with the poles of 
the proposed superprojectors.

This paper is organised as follows. Section 2 provides a brief review of the unitary representations of the AdS algebra $\mathfrak{so}(3, 2)$ and its $\cN=1$ supersymmetric extension $\mathfrak{osp}(1|4)$. We also review the structure of on-shell fields in AdS$_4$. 
Section 3 is devoted to the construction of  spin projection operators that map every unconstrained superfield on AdS$^{4|4}$ into one with the properties of a conserved current supermultiplet. On-shell supermultiplets in AdS$_4$ are studied in section 4. 
Section 5 is devoted to the component structure of the on-shell supermultiplets. In section 6 we present an off-shell model for the massive gravitino (superspin-1) multiplet. 
Factorisation of the superconformal higher-spin actions is described in section 7 and concluding comments are given in section \ref{Discussion}. 
The main body of the paper is accompanied by two appendices. Appendix A contains a list of identities that are indispensable for the derivation of many results in this paper. Appendix B 
is devoted to the derivation of partially massless gauge transformations, both in the non-supersymmetric and supersymmetric cases.

Throughout this work we will make use of various abbreviations to denote different types of irreducible superfields. In particular, a superfield $J_{\a(m)\ad(n)}$ satisfying \eqref{supercurrent} is simultaneously transverse linear and transverse anti-linear, and will be called TLAL. A superfield $J_{\a(m)}$ satisfying \eqref{supercurrent2} is simultaneously linear and transverse anti-linear, and will be called LTAL. Finally, a scalar superfield $J$ satisfying \eqref{supercurrent3} is simultaneously linear and anti-linear, and will be called LAL.


\section{Representations of the AdS (super)algebra} \label{section review}

In this section we collate the well-known facts concerning (i) the unitary representations 
of the AdS$_4$ algebra $\mathfrak{so}(3, 2)$ and its $\cN=1$ supersymmetric extension
$\mathfrak{osp}(1|4)$; and (ii)  the on-shell fields in AdS$_4$. 

\subsection{Unitary representations of $\mathfrak{so}(3, 2)$ and 
$\mathfrak{osp}(1|4)$} \label{section rep}

The irreducible unitary representations of the AdS${}_4$ algebra $\mathfrak{so}(3, 2)$ 
were studied in detail 
in~\cite{Dirac:1935zz, Dirac:1963ta, Fronsdal:1965zzb, Fronsdal:1974ew, Fronsdal:1975eq, Fronsdal:1975ac, Evans,Angelopoulos,AFFS}. 
The results of these studies were used in \cite{Heidenreich:1982rz}
to classify the irreducible unitary representations 
of the ${\cal N}=1$ AdS${}_4$ superalgebra $\mathfrak{osp} (1|4)$ 
 (see also~\cite{Nicolai:1984hb, deWit:1999ui} for a comprehensive review). 

The irreducible unitary representations of the AdS$_4$ algebra $\mathfrak{so}(3, 2)$ are specified by the lowest value $E_0$ of the energy $E$ and spin $s$,
and are traditionally denoted by $D (E_0, s)$.\footnote{The parameters $E_0$ and $s$ determine the values of the quadratic and quartic Casimir operators of  $\mathfrak{so}(3, 2)$, as shown in  \cite{Fronsdal:1974ew,Evans}. The energy $E$ is chosen to be dimensionless.
To restore dimensionful  energy, one has to rescale  $E \to |\mu| \,E$, where $\m\bar \m$ determines the AdS curvature \eqref{algebra}.}
The allowed  spin values
are $s = 0, 1/2, 1, \cdots $, the same as in Minkowski space. However, unlike in Minkowski space, unitarity imposes a bound on the allowed values of energy. According to the theorems proved in \cite{Evans,Angelopoulos}, $D(E_0, s)$ is unitary iff one of the following conditions holds: (i) $s=0$, $E_0 \geq \hf$; (ii) $s=\hf$,  $E_0\geq 1$, 
and (iii) $s\geq 1$, $E_0 \geq s+1$. 
The representations $D\big(\hf , 0\big) = {\rm Rac} $ and $D\big( 1, \hf \big) = {\rm Di} $ 
are known as the Dirac singletons \cite{Dirac:1963ta}.\footnote{It was found by Flato and Fronsdal \cite{Flato:1980zk,FF78} that the singletons are the square roots of massless particles in the sense that all two-singleton states are massless.} 
The representations 
$D(s+1, s) $ for $s>0$ and $D( 1, 0)  \oplus D (2, 0)$ are called massless since they contract  to the massless discrete helicity representations of the Poincar\'e group 
\cite{AFFS}. These representations prove to be restrictions of certain unitary representations of the conformal algebra   
$\mathfrak{so}(4, 2)$  to  $\mathfrak{so}(3, 2)$ \cite{AFFS,Barut:1970kp}. The remaining representations $D(E_0, s)$  are usually referred to as the massive AdS$_4$ representations. 

Irreducible unitary representations of the $\cN=1$ AdS$_4$ superalgebra  $\mathfrak{osp} (1|4)$, 
which we will call supermultiplets, 
are conveniently described by decomposing them into irreducible representations 
of $\mathfrak{so}(3, 2)$ in analogy 
with the case of the super Poincar\'e algebra in flat space. Each term in the decomposition is identified with a particle (or field)
carrying definite energy and spin. Unlike in flat space, particles in AdS${}_4$ belonging to the same supermultiplet do not carry the same energy.

There exist four types of unitary supermultiplets in AdS${}_4$.\footnote{Our notation for the AdS$_4$ supermultiplets follows the one introduced in \cite{Sibiryakov}. Fronsdal 
 \cite{Fronsdal:1985pc} used a different notation for these representations, 
$D^S(E_0, s)$.
} 
The first are known as massive representations and have the following decomposition
\begin{subequations} \label{E1}
\begin{align}
\mathfrak{S}\big({E}_0,s\big):=
D\big(E_0+ \frac{1}{2}, s- \frac{1}{2}\big)  \oplus &D\big( E_0, s\big) \oplus  D \big(E_0+1, s\big) \oplus  D\big(E_0+ \frac{1}{2}, s+ \frac{1}{2}\big)~, \label{E1.a} \\
 &\quad s>0\,, \quad E_0 >s+1\,, 
\end{align}
\end{subequations}
where the last inequality is the unitarity bound. This decomposition implies that a massive supermultiplet describes four particles in AdS${}_4$. 
When $s=0$ the first term in the right-hand side of \eqref{E1.a} is absent in the decomposition and the unitarity bound is also modified. 
The corresponding representations are referred to as the Wess-Zumino representations: 
\bea
\mathfrak{S}\big({E}_0,0\big):=D\big( E_0, 0 \big)  \oplus D \big(E_0+1, 0\big)
\oplus D\big(E_0+ \frac{1}{2},  \frac{1}{2}\big)  \,, \qquad E_0 >\frac{1}{2}\,. 
\label{E2}
\eea
When the unitarity bound is saturated, $E_0= s+1$ in eq.~\eqref{E1} or $E_0=\frac{1}{2}$ in eq.~\eqref{E2}, the supermultiplets get shortened. 
This yields  the other two types of representations. For $s>0$ the decomposition~\eqref{E1} contains fields of spin $s \geq 1$ and the shortening 
can be attributed to the appearance of  gauge invariance at the field theoretic level. The corresponding representations are called massless and are of the form 
\bea
\mathfrak{S}\big(s+1,s\big):=
D\big(s+1, s \big) \oplus  D\big(s+ \frac{3}{2}, s+ \frac{1}{2}\big) \,, \qquad s>0\,.
\label{E3}
\eea
This implies that a massless supermultiplet in AdS$_4$ consists of two physical component fields. Finally, the fourth type of representation occurs 
for $s=0, \ E_0= \frac{1}{2}$ and is called the Dirac supermultiplet 
\cite{Flato:1980zk,Fronsdal:1981gq}
(or super-singleton) 
\bea 
\mathfrak{S}\big( \hf, 0\big) = D\big(\hf , 0 \big) \oplus  D\big(1, \frac{1}{2}\big) \,.
\label{E4}
\eea
It unifies the bosonic and fermionic singletons.  

We will call the parameter $s$ in  \eqref{E1} and \eqref{E3}  the superspin. The Wess-Zumino  supermultiplets \eqref{E2} correspond to the superspin-0 representations.


\subsection{On-shell fields in AdS} \label{section2.2}

The geometry  of ${\rm AdS}_4$ can be described in terms of 
torsion-free Lorentz-covariant derivatives of the form 
\bea
\nabla_a = e_a + \o_a~, \qquad
e_a= e_a{}^m \partial_m  ~,
\eea
where $e_a{}^m $ is the inverse vielbein, and 
\bea
\o_a = \frac{1}{2}\,\o_a{}^{bc} M_{bc}
= \o_a{}^{\b \g} M_{\b \g}
+\bar \o_a{}^{\bd \gd} \bar M_{\bd \gd} ~,
\eea	
is the Lorentz connection. The Lorentz generators $M_{bc} \Leftrightarrow (M_{\b\g},{\bar M}_{\bd\gd})$ are defined in appendix \ref{appendixA}. 
The algebra of AdS$_4$ covariant derivatives is
 \begin{align}
 \big[\nabla_{\a\ad},\nabla_{\b\bd}\big]=-2\mu\mub\big(\ve_{\a\b}\bar{M}_{\ad\bd}+\ve_{\ad\bd}M_{\a\b}\big)
\quad \Longleftrightarrow \quad \big[\nabla_a,\nabla_b\big]=-\mu\mub M_{ab}
~,
 \label{algebra}
 \end{align} 
 where the parameter $\mu\mub > 0$ is related to the AdS$_4$ scalar curvature via $\mathcal{R}=-12\mu\mub$. The complex parameter $\m$ appears explicitly only in the algebra of AdS$^{4|4}$ covariant  derivatives given by eq.  \eqref{algebraS}.
 
 Below we make use of the quadratic Casimir of the AdS$_4$ group, whose realisation on tensor fields is
\begin{align}
\mathcal{Q}:=-\frac{1}{2}\nabla^{\a\ad}\nabla_{\a\ad}-\mu\mub \big(M^{\a\b}M_{\a\b}
+\bar{M}^{\ad\bd}\bar{M}_{\ad\bd}\big)~,\qquad \big[\mathcal{Q},\nabla_{\a\ad}\big]=0~. 
\label{Cas2}
\end{align}
One should keep in mind that $\nabla^a \nabla_a = -\frac{1}{2}\nabla^{\a\ad}\nabla_{\a\ad}$.

Given two positive integers $m$ and $n$, a tensor field $h_{\a(m)\ad(n)}$ of Lorentz type $(m/2,n/2)$ in AdS$_4$  is said to be on-shell if it satisfies the equations
\begin{subequations}\label{779}
\begin{align}
0~&=\big(\mathcal{Q}-\rho^2\big)h_{\a(m)\ad(n)}~,\label{779a}\\
0~&=\nabla^{\b\bd}h_{\a(m-1)\b\ad(n-1)\bd}~.\label{779b}
\end{align}
\end{subequations}
We say that such a field describes a spin $s=\frac{1}{2}(m+n)$ particle with pseudo-mass $\rho$.  
The following general result holds:
\bea
\rho^2 =\big[E_0(E_0-3) +s(s+1)\big]\mu\mub~,
\eea
see \cite{Nicolai:1984hb, deWit:1999ui} for pedagogical derivations.

As the name suggests, the pseudo-mass does not coincide with what is usually considered to be the physical mass $\rho_{\text{phys}}$. Rather, the two are related through
\begin{align}
\rho^2_{\text{phys}}=\rho^2-\tau_{(1,m,n)}\mu\mub~, \qquad
\tau_{(1,m,n)} = \hf \big( (m+n)^2 -4 \big) 
~,
\label{Mphys}
\end{align}
 where $\t_{(1,m,n)}$ is one of the partial massless parameters \eqref{PMval}. 
 The on-shell field $h_{\a(m)\ad(n)}$ corresponds to the irreducible representation $D\big(E_0,\frac{1}{2}(m+n)\big)$ where the minimal energy $E_0$ is related to the physical mass  through
\begin{align}
\rho_{\text{phys}}^2 
&=\big[E_0(E_0-3)-\frac{1}{4}(m+n+2)(m+n-4)\big]\mu\mub~. \label{Energy-Mass}
\end{align}
The unitarity bound for $m+n>1$
 is $E_0\geq  \frac{1}{2}(m+n+2)$, which in terms of the masses is
\begin{align}
\rho^2_{\text{phys}}\geq 0 \quad \implies \quad \rho^2 \geq \tau_{(1,m,n)}\mu\mub~, \qquad m+n>1~.
\label{Ubound}
\end{align}
With these relations in mind, we usually prefer to use the pseudo-mass as a representation label in place of $E_0$. As a caveat we note that there are two distinct values of $E_0$ leading to the same value of $\rho_{\text{phys}}^2$, 
\begin{align}
\big(E_0\big)_{\pm}=\frac{3}{2}\pm \frac{1}{2}
\sqrt{
4\frac{\rho^2_{\text{phys}}}{\mu\mub}+(m+n-1)^2
}
~.
\label{2.12}
\end{align}
However, the solution $(E_0)_-$ always violates the unitarity bound for $m+n>1$. Thus when referring to a unitary representation with $s >\hf$ and
pseudo-mass $\rho^2$, we are implicitly referring to the representation corresponding to $(E_0)_+$,
\begin{align}
E_0 =\frac{3}{2} + \frac{1}{2}
\sqrt{
4\frac{\rho^2_{\text{phys}}}{\mu\mub}+(m+n-1)^2}
~, \qquad m+n>1~.
\end{align}

For $s= 0, \hf $ (or, equivalently, $m+n=0,1$),  the unitary bound is $E_0\geq s +\hf$, 
and the solution  $(E_0)_-$ in \eqref{2.12} does not violate the unitarity bound for certain values of $\rho^2_{\text{phys}}$. For $s=0$ the values of $\rho^2_{\text{phys}}$ leading to $(E_0)_-\geq \hf$ are  restricted by
the condition
$- \frac 14 |\m|^2 \leq \rho^2_{\text{phys}} \leq  \frac 34 |\m|^2 $, which is known as  
the Breitenlohner-Freedman bound \cite{BreitenF}. 


\subsection{Partially massless and massive fields} \label{section NSPM}

Given two positive integers $m$ and $n$, the tensor field $h^{(t)}_{\a(m)\ad(n)}$ is said to be partially massless with depth-$t$ if it satisfies the on-shell conditions \eqref{779} such that its pseudo-mass takes the special value \cite{DeserW4, Zinoviev, Metsaev, KP20}
\begin{align}
\rho^2=\tau_{(t,m,n)}\mu\mub~,\qquad 1 \leq t \leq \text{min}(m,n)~, \label{PM0}
\end{align}
where the dimensionless constants $\tau_{(t,m,n)}$
are defined by 
\begin{align}
\tau_{(t,m,n)}:=\frac{1}{2}\Big((m+n-t+3)(m+n-t-1)+(t-1)(t+1)\Big) ~.
 \label{PMval}
\end{align} 
The specific feature of partially massless fields is that,
for a fixed $t$, the system of equations \eqref{779} and \eqref{PM0} admits a depth-$t$ gauge symmetry
\begin{align}
\delta_{\zeta}h^{(t)}_{\a(m)\ad(n)}=\nabla_{(\a_1(\ad_1}\cdots \nabla_{\a_t\ad_t}\zeta^{(t)}_{\a_{t+1}\dots\a_{m})\ad_{t+1}\dots\ad_{n})}~. \label{GTNSPM}
\end{align}
This is true as long as the gauge parameter $\zeta^{(t)}_{\a(m-t)\ad(n-t)}$ is also on-shell with the same pseudo-mass,
\begin{subequations}\label{GPP}
\begin{align}
0~&=\big(\mathcal{Q}-\tau_{(t,m,n)}\mu\mub\big)\zeta^{(t)}_{\a(m-t)\ad(n-t)}~,\label{GPPa}\\
0~&=\nabla^{\b\bd}\zeta^{(t)}_{\a(m-t-1)\b\ad(n-t-1)\bd}~.\label{GPPb}
\end{align}
\end{subequations}

It was demonstrated in \cite{KP20} that the parameters  $\tau_{(t,m,n)}$ defined by \eqref{PMval} determine the poles of the off-shell transverse projection operator for  fields
of Lorentz type $(m/2,n/2)$ in AdS$_4$. This observation leads to a new understanding of  the partially massless fields. 
Specifically, the gauge symmetry \eqref{GTNSPM} of the field $h^{(t)}_{\a(m)\ad(n)}$
 is associated with the pole $\rho^2=\tau_{(t,m,n)}\mu\mub$
of the corresponding spin projection operator in AdS$_4$.
In appendix \ref{appendixB} we provide a systematic derivation of this claim using the spin projection operators of \cite{KP20}. 

Strictly massless fields carry depth $t=1$ and therefore have mass given by 
\begin{align}
\rho^2=\tau_{(1,m,n)}\mu\mub \quad \implies \quad \rho_{\text{phys}}^2=0~.
\label{2.17}
\end{align}
This saturates the bound \eqref{Ubound} and hence defines a unitary representation of $\mathfrak{so}(3, 2)$. However, on account of the inequality 
\begin{align}
\tau_{(1,m,n)}> \tau_{(t,m,n)}~,\qquad 2 \leq t \leq \text{min}(m,n)~,
\end{align}
 the true partially massless representations 
are non-unitary. In particular, this means that there are two minimal energy values, 
\begin{align}
 (E_0)_{\pm}=\frac{3}{2}\pm\frac{1}{2}(m+n-2t+1)~,
 \end{align}
which are equally valid since they both violate the unitarity bound. In this work, whenever this ambiguity arises, we always implicitly choose the positive branch, $E_0\equiv (E_0)_+$.    
To distinguish the true partially-massless representations with depth $t$ and Lorentz type $(m/2,n/2)$, we will employ the notation  
\begin{align}
P\big(t,m,n\big)~,  \qquad2\leq t \leq \text{min}(m,n)~. 
 \label{NSPMrep}
\end{align}
Such a representation carries minimal energy $E_0=\frac{1}{2}(m+n)-t+2$.

 The massive representation of $\mathfrak{so}(3, 2)$ with spin $s=\frac{1}{2}(m+n)$
 is realised on the space of fields $h_{\a(m)\ad(n)}$  satisfying the equations \eqref{779}
 in which  $\r$ is constrained by 
 \begin{align}
 \rho^2>\tau_{(1,m,n)}\mu\mub \quad\implies\quad \rho_{\text{phys}}^2>0~,
 \end{align}
but is otherwise arbitrary. This restriction ensures the unitarity of the representation.


\section{AdS superprojectors} \label{main}

In this section we construct spin projection operators that map every unconstrained superfield on AdS$^{4|4}$ into one with the properties of a conserved current supermultiplet. 


\subsection{AdS superspace} \label{superspace}

Historically, the $\cN=1$ AdS superspace, AdS$^{4|4}$, was originally introduced 
\cite{Keck,Zumino77,IS} as the coset space
${\rm AdS^{4|4} } := {{\sOSp}(1|4)}/{{\sSO}(3,1)}$. 
The same superspace is equivalently realised as a unique maximally symmetric solution of 
the two off-shell formulations for $\cN=1$ AdS supergravity:
(i) the well-known minimal theory (see, e.g., \cite{GGRS,BK} for reviews); and 
(ii)  the more recently discovered non-minimal theory \cite{ButterK}. Here we prefer to use the second definition, since it allows us to obtain all information about the geometry of 
${\rm AdS}^{4|4}$ from the well-known supergravity results. 

As usual, we denote by  $z^M =(x^m , \q^\m, \bar \q_{\dot \m} ) $ the local coordinates
of ${\rm AdS}^{4|4}$. 
The geometry  of ${\rm AdS}^{4|4}$ is described in terms of covariant derivatives
of the form
\bea
\cD_A = (\cD_a , \cD_\a ,\bar \cD^\ad ) = E_A + \O_A~, \qquad
E_A = E_A{}^M \partial_M  ~,
\label{33.1}
\eea
where $E_A{}^M $ is the inverse superspace vielbein, and 
\bea
\O_A = \frac{1}{2}\,\O_A{}^{bc} M_{bc}
= \O_A{}^{\b \g} M_{\b \g}
+\bar \O_A{}^{\bd \gd} \bar M_{\bd \gd} ~,
\eea	
is the Lorentz connection. 
 The covariant derivatives 
obey the following graded commutation relations:
 \begin{subequations}\label{algebraS}
\bea
\{\cD_\a,\cDB_{\ad} \} &=& -2 \ri \cD_{\a\ad}~, \\
\{ \cD_\a, \cD_\b \} &=& -4\mub M_{\a\b}~, \qquad \{ \cDB_\ad, \cDB_\bd \} = 4 \m \bar{M}_{\ad \bd}~, \\
\, [\cD_\a , \cD_{\b\bd} ] &=& \ri \mub \varepsilon_{\a\b}\cDB_{\bd},  \qquad  \, [\cDB_\ad, \cD_{\b\bd} ] = - \ri \m \varepsilon_{\ad\bd}\cD_\b, \\
\, [\cD_{\a\ad},\cD_{\b\bd}] &=& - 2\mub \m (\varepsilon_{\a\b}\bar{M}_{\ad\bd} + \varepsilon_{\ad\bd}M_{\a\b} )~.
\eea
\end{subequations}
Here $\m\neq 0$ is a  complex parameter. We recall that $\m$ is related to the scalar curvature $\cR$ of  AdS$_4$ by the rule $\cR = -12 \m \bar \m$.

In what follows, we make extensive use of the quadratic Casimir of the AdS superalgebra, whose realisation on superfields takes the form \cite{BKS}
\begin{subequations} \label{Cas1}
\bea
\mathbb{Q}:=\Box+\frac{1}{4}\Big(\mu\mathcal{D}^2+\mub\bar{\mathcal{D}}^2\Big)
&-&\mu\mub\Big(M^{\a\b}M_{\a\b}+\bar{M}^{\ad\bd}\bar{M}_{\ad\bd}\Big)~, 
\quad 
\Box=\mathcal{D}^a\mathcal{D}_a
~,
\\
\big[\mathbb{Q},\mathcal{D}_{A}\big]&=&0~. 
\label{Cas1.b}
\eea
\end{subequations}
It is an instructive exercise to check, using the relations \eqref{algebraS},  that  \eqref{Cas1.b} holds.

As discussed in section \ref{section1}, 
 a complex tensor superfield 
 $ \G_{\a(m)\ad(n)}  $
 is said to be transverse linear if it satisfies the constraint \eqref{1.1},
whilst its conjugate obeys the constraint
\bea
\cD^\b \bar{\G}_{\b\a(n-1)\ad(m)} &=& 0 \quad \Longleftrightarrow \quad (\cD^2-2(n+2)\mub)\bar{\G}_{\a(n)\ad(m)} = 0  \label{TALCon}
\eea
and is called  transverse anti-linear.
Similarly, a complex superfield $G_{\a(m)\ad(n)}$ is said to be longitudinal linear if it satisfies the constraint \eqref{1.2},
whilst its conjugate is constrained by 
\bea
\cD_{(\a_1}\bar{G}_{\a_2\dots\a_{n+1})\ad(m)}&=&0 \quad \Longleftrightarrow \quad (\cD^2+2n\mub)\bar{G}_{\a(n)\ad(m)}=0~, \label{LALCon}
\eea
and is called longitudinal anti-linear.
We recall that for $n=0$ the constraint \eqref{1.2} defines a chiral superfield, 
	\bea \label{CLCon}
	\cDB_{\ad}G_{\a(m)}&=&0 \quad  \Longleftrightarrow \quad \cDB^2 G_{\a(m)}=0~.
	\eea
We also recall that for $n=0$ the first constraint in \eqref{1.1} is not defined, however 
one can consistently define complex linear superfields constrained by \eqref{1.3}. 

Given a complex tensor superfield ${\mathfrak V}_{\a(m)  \ad(n)} $ with $n \neq 0$, 
it can  be uniquely represented
as a sum of transverse linear and longitudinal linear multiplets, 
\bea
\mathfrak{V}_{\a(m) \ad(n)} = &-& 
\frac{1}{2 \mu (n+2)} \cDB^\gd \cDB_{(\gd} \mathfrak{V}_{\a(m) \ad_1 \dots  \ad_n)} 
- \frac{1}{2 \mu (n+1)} \cDB_{(\ad_1} \cDB^{|\gd|} \mathfrak{V}_{\a(m) \ad_2 \dots \ad_{n} ) \gd} 
~ . ~~~
 \label{UniqueDecom}
\eea
Choosing $\mathfrak{V}_{\a(m) \ad(n)} $ to be  transverse linear ($\G_{\a(m) \ad(n)} $)
or longitudinal linear ($G_{\a(m) \ad(n)} $), the above relation
gives
\begin{subequations}
\bea	\label{SolveTLCon}
\G_{\a(m)\ad(n)} &=& \cDB^\bd \x_{\a(m)\bd\ad(n)}~,\\
G_{\a(m)\ad(n)} &=& \cDB_{(\ad_1}\z_{\a(m)\ad_2...\ad_n)}~, \label{SolveLLCon}
\eea
\end{subequations}
for some prepotentials $ \x_{\a(m) \ad(n+1)}$ and $  \z_{\a(m) \ad(n-1)}$.
These relations provide  general solutions to the constraints \eqref{1.1}  and \eqref{1.2}.

Ref.  \cite{IS} introduced projectors $\mathcal{P}^{\perp }_n$ and 
$\mathcal{P}^{\parallel}_n$ 
which map the space of unconstrained superfields $\mathfrak{V}_{\a(m)\ad(n)}$ to the space of transverse linear \eqref{1.1} and longitudinal linear superfields \eqref{1.2}, 
respectively. These operators have the form
\begin{subequations}\label{TLLL}
	\bea
	\mathcal{P}^{\perp }_n &=& \frac{1}{4(n+1)\m} (\cDB^2+2n\m)~, \qquad \qquad \quad \cDB^\bd \mathcal{P}^{\perp }_n \mathfrak{V}_{\a(m)\bd\ad(n-1)} = 0~, \\
	\mathcal{P}^{\parallel}_n &=& -\frac{1}{4(n+1)\mu}(\cDB^2 - 2(n+2)\m)~, \qquad \cDB_{(\ad_1}\mathcal{P}^{\parallel}_n \mathfrak{V}_{\a(m)\ad_2 ... \ad_{n+1})} = 0 ~,
	\eea
\end{subequations}
and satisfy the projector properties
\be
\mathcal{P}^{\perp}_n \mathcal{P}^{\parallel}_n =0~,\qquad \mathcal{P}^{\parallel}_n \mathcal{P}^{\perp}_n  = 0~, \qquad \mathcal{P}^{\perp}_n + \mathcal{P}^{\parallel }_n =\mathds{1}~.
\ee
These properties imply 
that any superfield $\mathfrak{V}_{\a(m)\ad(n)}$ can be uniquely represented as a sum of transverse linear and longitudinal linear superfields,
\bea
\mathfrak{V}_{\a(m)\ad(n)} = \G_{\a(m)\ad(n)} + G_{\a(m)\ad(n)}~,
\eea
which agrees with \eqref{UniqueDecom}. It should be pointed out that the projectors 
\eqref{TLLL} are non-analytic in $\m$. 

Finally, let us introduce the following operators:
\begin{subequations} \label{Sproj}
\begin{align}
\mathcal{P}_{(0)}&:=-\frac{1}{8\mathbb{Q}}\mathcal{D}^{\b}\big(\bar{\mathcal{D}}^2-4\mu\big)\mathcal{D}_{\b}~,\\
 \mathcal{P}_{(+)}&:=\frac{1}{16\mathbb{Q}}\big(\bar{\mathcal{D}}^2-4\mu\big)\mathcal{D}^2~,\\ 
 \mathcal{P}_{(-)}&:=\frac{1}{16\mathbb{Q}}\big(\mathcal{D}^2-4\mub\big)\bar{\mathcal{D}}^2~.
\end{align}
\end{subequations}
By making use of the identities in appendix \ref{appendixA}, one may show that when restricted to the space of scalar superfields they satisfy the projector properties 
\begin{align}
\mathds{1}=\sum_{i}\mathcal{P}_{(i)}~,\qquad \mathcal{P}_{(i)}\mathcal{P}_{(j)}=\delta_{ij}\mathcal{P}_{(j)}~, \label{projprop}
\end{align}
for $i=0,+,-$. Here $ \mathcal{P}_{(+)}$ and $ \mathcal{P}_{(-)}$ are the chiral and antichiral projectors, respectively, while $ \mathcal{P}_{(0)}$ projects onto  the space of LAL superfields. 
Using these projectors, it is always possible to decompose an unconstrained complex scalar superfield $\mathfrak{V}$  as
\begin{align}
\mathfrak{V}=\mathfrak{L}+\sigma+\bar{\rho}~,
\end{align}
where the complex superfield $\mathfrak{L}$ is simultaneously linear and anti-linear, $\sigma$ is chiral and $\bar{\rho}$ is anti-chiral
\begin{align}
\big(\mathcal{D}^2-4\mub\big)\mathfrak{L}=\big(\bar{\mathcal{D}}^2-4\mu\big)\mathfrak{L} =0~,\qquad \bar{\mathcal{D}}_{\ad}\s=0~,\qquad \mathcal{D}_{\a}\bar{\rho}=0~.
\end{align}
If $\mathfrak{V}$ is real, then $\mathfrak{L}$ is also real and $\bar{\rho}=\bar{\s}$.
In the flat superspace limit, the projectors reduce to those constructed by Salam and Strathdee \cite{SalamS}.


\subsection{Construction of superspin projection operators} \label{construction}

Our construction of AdS$_4$ superprojectors will be based on certain properties of the following theories: (i) the massless supersymmetric higher-spin gauge models in AdS${}_4$ \cite{KS94}; 
and (ii) the superconformal higher-spin (SCHS) gauge theories in AdS${}_4$  \cite{KMT,KP19,KR}.

Let $s$ be a positive integer. We recall that there are two dually equivalent gauge formulations for the massless superspin-$(s+\hf)$ multiplet \cite{KS94}. 
They both involve the same superconformal prepotential, 
$H_{\a(s)\ad(s)}$,
 but different compensating multiplets. The prepotential 
$H_{\a(s)\ad(s)} $ is real and possesses the gauge freedom 
\bea
\delta_{\zeta}H_{\a(s)\ad(s)}=\bar{\mathcal{D}}_{(\ad_1}\z_{\a(s)\ad_2\dots\ad_s)}
-\mathcal{D}_{(\a_1}\bar \z_{\a_2\dots\a_s)\ad(s)}~,
\eea
with the gauge parameter $\z_{\a(s)\ad (s-1)}$ being unconstrained. Associated with 
$H_{\a(s) \ad(s)}$ is the gauge-invariant chiral field strength
\bea
\mathfrak{W}_{\a (2s+1)} (H) &=& -\frac{1}{4}(\bar \cD^2 -4\m)\cD_{(\a_1}{}^{\bd_1} \cdots \cD_{\a_s}{}^{\bd_s}
\cD_{\a_{s+1}} H_{\a_{s+2} \dots \a_{2s+1} )\bd_1 \dots \bd_s} ~.
\label{3.16}
\eea
The field strength $\mathfrak{W}_{\a (2s+1)} (H) $ and its conjugate are the only independent gauge-invariant field strengths which survive on the mass shell. 

The massless superspin-$s$ multiplet can be described using a superconformal
 prepotential $\J_{\a(s)\ad(s-1)} $,  
 its conjugate $\bar \J_{\a(s-1)\ad(s)}$ and certain compensating multiplets  \cite{KS94,BHK}.
The gauge transformation law of $\J_{\a(s) \ad(s-1)} $ is 
\bea
\delta_{\zeta,\xi}\J_{\a(s)\ad(s-1)}=\bar{\mathcal{D}}_{(\ad_1}\zeta_{\a(s)\ad_2\dots\ad_{s-1})}+\mathcal{D}_{(\a_1}\xi_{\a_2\dots\a_s)\ad(s-1)}~,
\eea
with the gauge parameters $\z_{\a(s) \ad(s-2)}$ and $\x_{\a(s-1) \ad(s-1)}$ being unconstrained. 
 Associated with 
$\J_{\a(s) \ad(s-1)}$ is the gauge-invariant chiral field strength
\bea
\mathfrak{W}_{\a (2s)} (\J) &=& -\frac{1}{4}(\bar \cD^2 -4\m)\cD_{(\a_1}{}^{\bd_1} \cdots \cD_{\a_{s-1}}{}^{\bd_{s-1}}
\cD_{\a_{s}} \J_{\a_{s+1} \dots \a_{2s} )\bd_1 \dots \bd_{s-1}} ~.
\label{3.18}
\eea
The field strength $\mathfrak{W}_{\a (2s)} (\J) $ and its conjugate prove to be the only independent gauge-invariant field strengths which survive on the mass shell.\footnote{Off the mass shell, there exists one more gauge-invariant chiral field strength, 
$\mathfrak{W}_{\a (2s)} (\bar \J) $, defined according to the general rule \eqref{HSW2}.
However,  it may be shown that 
$\mathfrak{W}_{\a (2s)} (\bar \J) \propto \mathfrak{W}_{\a (2s)} (\J) $ on the mass shell.
In the flat superspace limit, $\mathfrak{W}_{\a (2s)} (\bar \J) =0$ on-shell. 
} 

The gauge prepotentials $H_{\a(s) \ad(s)}$ and $\J_{\a(s) \ad(s-1)}$ 
and the corresponding gauge-invariant field strengths \eqref{3.16} and \eqref{3.18}
were used to construct superconformal higher-spin theories  \cite{KMT,KP19}.
More general superconformal higher-spin models in AdS$_4$
can be introduced
 by making use of  gauge prepotentials $\Phi_{\a(m)\ad(n)}$, 
 with $m$ and $n$ positive integers.\footnote{For the superconfomal gauge multiplets with either $m=0$ or $n=0$ see \cite{KR}.}
  The superfield $\Phi_{\a(m)\ad(n)}$ is defined 
modulo the gauge transformations \cite{KMT,KP19}
\begin{align}
\delta_{\zeta,\xi}\Phi_{\a(m)\ad(n)}=\bar{\mathcal{D}}_{(\ad_1}\zeta_{\a(m)\ad_2\dots\ad_n)}+\mathcal{D}_{(\a_1}\xi_{\a_2\dots\a_m)\ad(n)}~, \label{GT}
\end{align}
with the gauge parameters being unconstrained.
Associated with  $\Phi_{\a(m)\ad(n)}$ and its complex conjugate  $\bar \Phi_{\a(n)\ad(m)}$
are the  chiral field strengths (also known as  the linearised higher-spin super-Weyl tensors) 
\begin{subequations}
\begin{align}
\mathfrak{W}_{\a(m+n+1)}(\Phi)&= -\frac{1}{4}\big(\bar{\mathcal{D}}^2-4\mu\big)\mathcal{D}_{(\a_1}{}^{\bd_1}\cdots\mathcal{D}_{\a_n}{}^{\bd_n}\mathcal{D}_{\a_{n+1}}\Phi_{\a_{n+2}\dots\a_{m+n+1})\bd(n)}~, \label{HSW1}\\
\mathfrak{W}_{\a(m+n+1)}(\bar{\Phi})&= -\frac{1}{4}\big(\bar{\mathcal{D}}^2-4\mu\big)\mathcal{D}_{(\a_1}{}^{\bd_1}\cdots\mathcal{D}_{\a_m}{}^{\bd_m}\mathcal{D}_{\a_{m+1}}\bar{\Phi}_{\a_{m+2}\dots\a_{m+n+1})\bd(m)}~,\label{HSW2}
\end{align}
\end{subequations}
which are  invariant under the gauge transformations \eqref{GT}
\begin{align}
\delta_{\zeta,\xi}\mathfrak{W}_{\a(m+n+1)}(\Phi)=0~,\qquad \delta_{\zeta,\xi}\mathfrak{W}_{\a(m+n+1)}(\bar{\Phi})=0~.
\end{align}
The gauge-invariant action $S_{\text{SCHS}}^{(m,n)}[\Phi,\bar{\Phi}]$, 
which describes the dynamics of $\Phi_{\a(m)\ad(n)}$ and its conjugate  $\bar \Phi_{\a(n)\ad(m)}$,
is typically written as a functional over the chiral subspace 
of the full superspace
\cite{KMT,KP19}.
The specific feature of AdS${}^{4|4}$ is the identity \cite{Siegel78}
\bea
 \int\rd^4x\rd^2\q\, \cE \,\cL_{\rm c}
= \int\rd^4x\rd^2\q\rd^2\qb\, \frac{E}{\m} \,\cL_{\rm c}~, \qquad \bar \cD_\ad \cL_{\rm c} =0~,
\label{full-chiral}
\eea
which relates the integration over the chiral subspace to that over the full superspace;
here $\cE$ denotes the chiral integration measure  and $E^{-1} = {\rm Ber} (E_A{}^M)$. 
Keeping in mind  \eqref{full-chiral}, 
the gauge-invariant action $S_{\text{SCHS}}^{(m,n)}[\Phi,\bar{\Phi}]$ is given by 
\begin{align}
S_{\text{SCHS}}^{(m,n)}[\Phi,\bar{\Phi}]=\frac{1}{2}\text{i}^{m+n}\int 
\text{d}^{4|4}z \, \frac{E}{\m} \, 
 \mathfrak W^{\a(m+n+1)}(\Phi)\mathfrak{W}_{\a(m+n+1)}(\bar{\Phi}) +{\rm c.c.}
 ~, \label{SCHSchiral}
\end{align}
where we have denoted $\rd^{4|4}z = \rd^4x\rd^2\q\rd^2\qb$.   
Upon integrating by parts, the action \eqref{SCHSchiral} may be written in the suggestive forms
\begin{subequations}
\begin{align}
S_{\text{SCHS}}^{(m,n)}[\Phi,\bar{\Phi}]&=\text{i}^{m+n}\int \text{d}^{4|4}z \, E \, \bar{\Phi}^{\a(n)\ad(m)}\mathfrak{B}_{\a(n)\ad(m)}(\Phi) +\text{c.c. } \label{SCHS1}\\
&=\text{i}^{m+n}\int \text{d}^{4|4}z \, E \, \bar{\Phi}^{\a(n)\ad(m)}\widehat{\mathfrak{B}}_{\a(n)\ad(m)}(\Phi) +\rm{c.c.}\label{SCHS2}
\end{align}
\end{subequations}
They are suggestive because, in addition to being gauge invariant, the linearised higher-spin super-Bach tensors\footnote{Proving the equivalence of \eqref{Bach1} and \eqref{Bach2}, and hence \eqref{SCHS1} and \eqref{SCHS2}, is non-trivial and is closely related to the coincidence relation \eqref{coincidence}.}
 \cite{KPR}
\begin{subequations} \label{Bach}
\begin{align}
\mathfrak{B}_{\a(n)\ad(m)}(\Phi)&= \frac{1}{2}\mathcal{D}_{(\ad_1}{}^{\b_1}\cdots\mathcal{D}_{\ad_{m})}{}^{\b_m}\mathcal{D}^{\b_{m+1}}\mathfrak{W}_{\a(n)\b(m+1)}(\Phi)~,\label{Bach1}\\
\widehat{\mathfrak{B}}_{\a(n)\ad(m)}(\Phi)&=\frac{1}{2}\mathcal{D}_{(\a_1}{}^{\bd_1}\cdots\mathcal{D}_{\a_{n})}{}^{\bd_n}\bar{\mathcal{D}}^{\bd_{n+1}}\overline{\mathfrak{W}}_{\ad(m)\bd(n+1)}(\Phi)~, \label{Bach2}
\end{align}
\end{subequations}
are simultaneously transverse linear and transverse anti-linear (TLAL)
\begin{subequations}
\begin{align}
\mathcal{D}^{\b}\mathfrak{B}_{\b\a(n-1)\ad(m)}(\Phi)&= 0~, \qquad \bar{\mathcal{D}}^{\bd}\mathfrak{B}_{\a(n)\ad(m-1)\bd}(\Phi)=0~,\\
\mathcal{D}^{\b}\widehat{\mathfrak{B}}_{\b\a(n-1)\ad(m)}(\Phi)&= 0~, \qquad \bar{\mathcal{D}}^{\bd}\widehat{\mathfrak{B}}_{\a(n)\ad(m-1)\bd}(\Phi)=0~.
\end{align}
\end{subequations}

Our ansatz for the TLAL projectors is based on the higher-derivative descendants \eqref{Bach1} and \eqref{Bach2}, since the latter are each TLAL.  However, it is clear that any projector must preserve the rank of the tensor superfield on which it acts. This can be achieved by appropriately removing or inserting vector derivatives in \eqref{Bach} to convert the indices. With these remarks in mind, for positive integers $m$ and $n$, we define the following two differential operators by their action on an unconstrained superfield $\Phi_{\a(m)\ad(n)}$ 
\begin{subequations}\label{strippedprojectors}
\begin{align}
\mathbb{P}_{\a(m)\ad(n)}(\Phi)&=-\frac{1}{8}\cD_{(\ad_1}{}^{\b_1}\dots \cD_{\ad_n)}{}^{\b_n}\cD^{\g}(\cDB^2-4\mu  )\cD_{(\b_1}{}^{\bd_1}\dots\cD_{\b_n}{}^{\bd_n}\cD_\g \Phi_{\a_1\dots\a_m)\bd(n)}~,\label{P1}\\
\widehat{\mathbb{P}}_{\a(m)\ad(n)}(\Phi)&=\frac{1}{8}\cD_{(\a_1}{}^{\bd_1}\dots \cD_{\a_m)}{}^{\bd_m}\cDB^{\gd} (\cD^2-4\mub  )\cD_{(\bd_1}{}^{\b_1}\dots\cD_{\bd_m}{}^{\b_m}\cDB_\gd \Phi_{\b(m)\ad_1\dots\ad_n)}~.\label{P2}
\end{align}
\end{subequations}
From here one can define the operators
\begin{subequations}\label{projectors}
\begin{align}
\Pi^{(m,n)}\Phi_{\a(m)\ad(n)}\equiv \P_{\a(m)\ad(n)}(\Phi)&:=\bigg [\prod_{t=1}^{n+1} \big ( \mathbb{Q}-\l_{(t,m,n)}\m \mub \big ) \bigg ]^{-1}\mathbb{P}_{\a(m)\ad(n)}(\Phi)~, \label{proj1}\\
\widehat{\Pi}^{(m,n)}\Phi_{\a(m)\ad(n)}\equiv \widehat{\P}_{\a(m)\ad(n)}(\Phi)&:=\bigg [ \prod_{t=1}^{m+1} \big ( \mathbb{Q}-\l_{(t,m,n)}\m \mub \big ) \bigg ]^{-1} \widehat{\mathbb{P}}_{\a(m)\ad(n)}(\Phi)~, \label{proj2}
\end{align}
where the $\lambda_{(t,m,n)}$ are dimensionless constants defined by
\begin{align}
\l_{(t,m,n)} &= \hf \Big ( (m+n-t+1)(m+n-t+4) + t(t-1) \Big )~.
\label{SPM}
\end{align}
\end{subequations}
	The operators \eqref{projectors} project onto the space of TLAL superfields
\begin{subequations}\label{TLAproj}
\begin{align}
\cDB^\bd \P_{\a(m)\ad(n-1)\bd}(\Phi)&=0 ~,\qquad  \cDB^\bd \widehat{\P}_{\a(m)\ad(n-1)\bd}(\Phi)=0~,\\
\cD^\b \P_{\b\a(m-1)\ad(n)}(\Phi)&=0 ~,\qquad \cD^\b \widehat{\P}_{\b\a(m-1)\ad(n)}(\Phi)=0~.
\end{align}
\end{subequations}
 The operators \eqref{strippedprojectors} also satisfy \eqref{TLAproj}, but it is only the operators \eqref{projectors} which square to themselves
\begin{subequations}
\begin{align}
\P^{(m,n)}\P^{(m,n)}\Phi_{\a(m)\ad(n)} &= \P^{(m,n)}\Phi_{\a(m)\ad(n)}~,\\
\widehat{\P}^{(m,n)}\widehat{\P}^{(m,n)}\Phi_{\a(m)\ad(n)} &= \widehat{\P}^{(m,n)}\Phi_{\a(m)\ad(n)}~,
\end{align}
\end{subequations}
and are hence TLAL projectors. Furthermore, one can consider (say) $\Pi^{(m,n)}$ to be the unique TLAL projector, since it can be shown that the two types of projectors \eqref{proj1} and \eqref{proj2} actually coincide
\begin{align}
\P_{\a(m)\ad(n)}(\Phi) = \widehat{\P}_{\a(m)\ad(n)}(\Phi) ~. \label{coincidence}
\end{align}

In the case when $m>n=0$, the operators analogous to \eqref{strippedprojectors} take the form
\begin{subequations}
\begin{align}
\mathbb{P}_{\a(m)}(\Phi)&=-\frac{1}{8}\cD^{\g}(\cDB^2-4\mu  )\cD_{(\g} \Phi_{\a_1\dots\a_m)}~,\\
\widehat{\mathbb{P}}_{\a(m)}(\Phi)&=\frac{1}{8}\cD_{(\a_1}{}^{\bd_1}\dots \cD_{\a_m)}{}^{\bd_m}\cDB^{\gd} (\cD^2-4\mub  )\cD_{(\bd_1}{}^{\b_1}\dots\cD_{\bd_m}{}^{\b_m}\cDB_{\gd)} \Phi_{\b(m)}~,
\end{align}
\end{subequations}
whilst the projectors $\Pi^{(m)} :=\Pi^{(m,0)} $ and  $\widehat \Pi^{(m)} := \widehat \Pi^{(m,0)} $ 
are
\begin{subequations} \label{projectors-m}
\begin{align}
\Pi^{(m)}\Phi_{\a(m)}\equiv \P_{\a(m)}(\Phi)&:= \Big ( \mathbb{Q}-\l_{(1,m,0)}\m \mub \Big )^{-1}\mathbb{P}_{\a(m)}(\Phi)~, \label{3.33a} \\
\widehat{\Pi}^{(m)}\Phi_{\a(m)}\equiv \widehat{\P}_{\a(m)}(\Phi)&:=\bigg [ \prod_{t=1}^{m+1} \big ( \mathbb{Q}-\l_{(t,m,0)}\m \mub \big ) \bigg ]^{-1} \widehat{\mathbb{P}}_{\a(m)}(\Phi)~. 
\end{align}
\end{subequations}
These operators square to themselves but, in contrast to \eqref{TLAproj}, they project onto the subspace of simultaneously linear and transverse anti-linear (LTAL) superfields,
\begin{subequations} \label{3.34}
\begin{align}
\mathcal{D}^{\b}\Pi_{\a(m-1)\b}(\Phi)&=0~, \qquad ~~~~\phantom{..}\mathcal{D}^{\b}\widehat{\Pi}_{\a(m-1)\b}(\Phi)=0~, \\
\big(\bar{\mathcal{D}}^2-4\mu\big)\Pi_{\a(m)}(\Phi)&=0~, \qquad \big(\bar{\mathcal{D}}^2-4\mu\big)\widehat{\Pi}_{\a(m)}(\Phi)=0~.~~~~~~~
\end{align}
\end{subequations}
Once again one may show that the two types of projectors coincide, $\Pi_{\a(m)}(\Phi)=\widehat{\Pi}_{\a(m)}(\Phi)$. In the case when $n>m=0$, the corresponding projectors $\overline{\Pi}_{\ad(n)}(\bar \Phi)$ and $\widehat{\overline{\Pi}}_{\ad(n)}(\bar \Phi)$ can be obtained by complex conjugation and similar comments apply. This time however they project onto the subspace of simultaneously anti-linear and transverse linear superfields. 

When both $m\neq 0$ and $n\neq 0$, it may be shown that the orthogonal complement acting on $\Phi_{\a(m)\ad(n)}$ can be expressed as the following sum
\be
\Pi_{~||}^{(m,n)}\Phi_{\a(m)\ad(n)}:=\big(\mathds{1}-\P^{(m,n)}\big)\Phi_{\a(m)\ad(n)} = \bar{\mathcal{D}}_{(\ad_1}\Psi_{\a(m)\ad_2\dots\ad_{n})} + \mathcal{D}_{(\a_1}\O_{\a_2\dots\a_m)\ad(n)}
\ee
for some unconstrained superfields $\Psi_{\a(m)\ad(n-1)}$ and $\O_{\a(m-1)\ad(n)}$. We see that $\Pi_{~||}^{(m,n)}$ projects $\Phi_{\a(m)\ad(n)}$ onto the union of spaces of longitudinal linear and longitudinal anti-linear superfields. 

If instead $m>0$ and $n=0$, the orthogonal complement may be written as
\begin{align}
\Pi_{~||}^{(m)}\Phi_{\a(m)}:=\big(\mathds{1}-\P^{(m)}\big)\Phi_{\a(m)} = \mathcal{D}_{(\a_1}\O_{\a_2\dots\a_m)}+\Lambda_{\a(m)}~, \qquad 
\bar{\mathcal{D}}_{\bd}\Lambda_{\a(m)}=0~,
\end{align}
for some unconstrained 
$\O_{\a(m-1)}$ and chiral   
$\Lambda_{\a(m)}$ superfields. 
On the otherhand, if $n>0$ and $m=0$,  then the orthogonal complement splits into the sum of a longitudinal linear superfield and an anti-chiral superfield,
\begin{align}
\overline{\Pi}_{~||}^{(n)}\bar{\Phi}_{\ad(n)}:=\big(\mathds{1}-\overline{\P}^{(n)}\big)\bar{\Phi}_{\ad(n)} = \bar{\mathcal{D}}_{(\ad_1}\bar{\Psi}_{\ad_2\dots\ad_n)}+\bar{\Lambda}_{\ad(n)}~, \qquad \mathcal{D}_{\b}\bar{\Lambda}_{\ad(n)}=0~.
\end{align}

As the first application of the superprojectors, 
we provide new expressions for the superconformal actions \eqref{SCHSchiral}.
For the higher-spin gauge supermultiplets  $H_{\a(s)\ad(s)}$ and $\Psi_{\a(s)\ad(s-1)}$
they take the simple and elegant forms
\begin{subequations}\label{3.40}
\begin{align}
S_{\text{SCHS}}^{(s,s)}[H]&=2(-1)^s\int \text{d}^{4|4}z\, E \, H^{\a(s)\ad(s)}\prod_{t=1}^{s+1} \big ( \mathbb{Q}-\l_{(t,s,s)}\m \mub \big )\Pi^{(s,s)} H_{\a(s)\ad(s)}~,\\
S_{\text{SCHS}}^{(s,s-1)}[\Psi, \bar{\Psi}]&=\text{i}(-1)^{s+1}\int \text{d}^{4|4}z\, E \, \bar{\Psi}^{\a(s-1)\ad(s)}\prod_{t=1}^{s} \big ( \mathbb{Q}-\l_{(t,s,s-1)}\m \mub \big )\notag\\
&~~~~~~~~~~~~~~~~~~~~~~~~~~~~~~~~~\times \mathcal{D}_{\ad}{}^{\b}\Pi^{(s,s-1)} \Psi_{\b\a(s-1)\ad(s-1)}+\text{c.c.}
\end{align}
\end{subequations}
These expressions are analogous to those provided for the conformal 
higher-spin models in Minkowski space  \cite{FT}
and in AdS$_4$ \cite{KP20}. 


\subsection{Decomposing unconstrained superfields into irreducible parts} \label{decomp}

In section \ref{construction} it was shown that the operator $\Pi^{(m,n)}$ projects every unconstrained superfield $\Phi_{\a(m)\ad(n)}$ to the space of
TLAL superfields, whilst its complement $\Pi_{~||}^{(m,n)}$ projects onto the union of the spaces of longitudinal linear and anti-linear superfields. This allows us to decompose $\Phi_{\a(m)\ad(n)}$ as follows
\begin{align}
\Phi_{\a(m)\ad(n)}=\phi_{\a(m)\ad(n)}+\bar{\mathcal{D}}_{(\ad_1}\z_{\a(m)\ad_2\dots\ad_n)}+\mathcal{D}_{(\a_1}\x_{\a_2\dots\a_m)\ad(n)}~, \label{decomp0}
\end{align}
where $\phi_{\a(m)\ad(n)}$ is TLAL and is hence irreducible, whilst $\z_{\a(m)\ad(n-1)}$ and $\x_{\a(m-1)\ad(n)}$ are unconstrained and thus reducible. One can then repeat this decomposition on the lower-rank unconstrained superfields in \eqref{decomp0}. After reiterating this procedure a finite number of times, one eventually arrives at a decomposition of $\Phi_{\a(m)\ad(n)}$ solely in terms of irreducible superfields. 
In presenting the resulting decompositions, we will use the notational convention introduced in \cite{Vasiliev} and adopted in \cite{KS94}:
\bea
V_{ \a(n)} U_{\a(m) }&=& V_{ (\a_1 \dots \a_n} U_{\a_{n+1} \dots \a_{n+m})}~. \label{symmnot}
\eea
 
In the $m>n$ case,  the result of this process is
\begin{align}
\Phi_{\a(m)\ad(n)}&=\sum_{t=0}^{n} (\mathcal{D}_{\a\ad})^t 
\phi_{\a (m-t) \ad (n-t)} 
+\sum_{t=0}^{n-1}\big[\mathcal{D}_{\a},\bar{\mathcal{D}}_{\ad}\big]
(\mathcal{D}_{\a\ad})^t
\psi_{\a (m-t-1) \ad (n-t-1) }\non\\
&+\sum_{t=0}^{n} (\mathcal{D}_{\a \ad} )^t \mathcal{D}_{\a}
\chi_{\a (m-t-1 ) \ad(n-t) } 
+\sum_{t=0}^{n-1} (\mathcal{D}_{\a \ad} )^t  \bar{\mathcal{D}}_{\ad} 
\varphi_{\a (m-t) \ad (n-t-1) } \non\\
&+(\mathcal{D}_{\a \ad} )^n
\Big(\s_{\a (m-n)} + \mathcal{D}_{\a }\s_{\a(m-n-1) }\Big) ~,
\label{decomp1}
\end{align}
for some irreducible complex superfields $\phi, \psi, \chi$ and $\varphi$ whose properties are summarised in table \ref{table 1}. The superfields
$\s_{\a (m-n)} $ and $\s_{\a(m-n-1) }$ in \eqref{decomp1} are chiral, 
\begin{align}
\bar{\mathcal{D}}_{\ad}\s_{\a(m-n)}=0~,\qquad \bar{\mathcal{D}}_{\ad}\s_{\a(m-n-1)}=0~.
\end{align}

\begin{table}[h]
\begin{center}
\begin{tabular}{|c|c|c|c|}
\hline
~  & $0\leq t \leq n-2$           & $t=n-1$ & $t=n$  \Tstrut\Bstrut\\ \hline
  $\phi_{\a(m-t) \ad(n-t)}$       &  TLAL   & TLAL  & LTAL\Tstrut\\ 
  $\psi_{\a(m-t-1) \ad(n-t-1)}$   &  TLAL   & TLAL  & --\\ 
  $\chi_{\a(m-t-1) \ad(n-t)}$     &  TLAL   & TLAL  & LTAL \\ 
  $\varphi_{\a(m-t) \ad(n-t-1)}$  &  TLAL   & LTAL   & -- \\
     \hline
\end{tabular}
\end{center} 
\vspace{-15pt} 
\caption{Properties of the superfields appearing in \eqref{decomp1}.}
\label{table 1}
\end{table}

If instead $m=n=s$, then we may further impose the reality condition 
\begin{align}
H_{\a(s)\ad(s)}:=\Phi_{\a(s)\ad(s)}=\bar{H}_{\a(s)\ad(s)}~. \label{reality}
\end{align}
In contrast to \eqref{decomp1}, $H_{\a(s)\ad(s)}$ now decomposes as
\begin{align}
H_{\a(s)\ad(s)}&=\sum_{t=0}^{s} (\mathcal{D}_{\a \ad} )^t
\phi_{\a (s-t)\ad (s-t)} 
+\sum_{t=0}^{s-1}\big[\mathcal{D}_{\a},\bar{\mathcal{D}}_{\ad}\big]
(\mathcal{D}_{\a\ad} )^t
\psi_{\a (s-t-1 )\ad (s-t-1)}\non\\
&+\sum_{t=0}^{s-1} (\mathcal{D}_{\a\ad})^t
\Big(\bar{\mathcal{D}}_{\ad}\chi_{\a( s-t)\ad (s-t-1)}+\text{c.c.}\Big) 
+ (\mathcal{D}_{\a\ad})^s
\big(\s+\bar{\s}\big)~ ,\label{decomp2}
\end{align}
for some irreducible complex superfield $\chi_{\a(s-t) \ad(s-t-1)}$ and 
irreducible real superfields
\begin{align} 
\phi_{\a(s-t)\ad(s-t)}=\bar{\phi}_{\a(s-t)\ad(s-t)}~, \qquad \psi_{\a(s-t-1)\ad(s-t-1)}=\bar{\psi}_{\a(s-t-1)\ad(s-t-1)}~,
\end{align}
whose properties are described in table \ref{table 2}. 
The scalar $\s$ in \eqref{decomp2} is chiral, 
$\bar{\mathcal{D}}_{\ad}\s=0$.
\begin{table}[h]
\begin{center}
\begin{tabular}{|c|c|c|c|}
\hline
~  & $0\leq t \leq s-2$     & $t=s-1$ & $t=s$  \Tstrut\Bstrut\\ \hline
  $\phi_{\a(s-t) \ad(s-t)}$        &  TLAL   & TLAL  & LAL \Tstrut\\ 
  $\psi_{\a(s-t-1) \ad(s-t-1)}$    &  TLAL   & LAL     & --\\ 
  $\chi_{\a(s-t) \ad(s-t-1)}$      &  TLAL   & LTAL   & -- \\ 
     \hline
\end{tabular}
\end{center} 
\vspace{-15pt} 
\caption{Properties of the superfields appearing in \eqref{decomp2}.}
\label{table 2}
\end{table}

The above decompositions were used, albeit without derivation,  
in \cite{BKS} for the covariant quantisation of 
 the massless supersymmetric higher-spin models in AdS${}_4$ \cite{KS94}.


\section{On-shell supermultiplets in AdS} \label{section 3}

In this section we confine our attention to on-shell superfields $\phi_{\a(m)\ad(n)}$. When $m\neq 0$ and $n\neq 0$ we define these to satisfy the mass-shell equation
\begin{subequations}\label{OS}
\bea
 \big(\mathbb{Q}-M^2\big)\phi_{\a(m)\ad(n)}=0~,\label{OS1}
 \eea
 and the irreducibility conditions
 \bea
 \mathcal{D}^{\b}\phi_{\b\a(m-1)\ad(n)}=0~,\qquad \bar{\mathcal{D}}^{\bd}\phi_{\a(m)\ad(n-1)\bd}=0 ~. \label{OS2}
\eea
\end{subequations}
Hence the superfield $\phi_{\a(m)\ad(n)}$ is simultaneously transverse linear and transverse anti-linear (TLAL), and is said to have super-spin $s=\frac{1}{2}(m+n+1)$ and pseudo-mass $M$. In the case when $m>n=0$ and $n>m=0$, the condition \eqref{OS2} should be modified to
\begin{subequations} \label{OS3}
\begin{align}
\mathcal{D}^{\b}\phi_{\a(m-1)\b}&=0~, \qquad \big(\bar{\mathcal{D}}^2-4\mu\big)\phi_{\a(m)}=0~,  \label{OS3.a}\\
\bar{\mathcal{D}}^{\bd}\bar{\phi}_{\ad(n-1)\bd}&=0~, \qquad \big(\mathcal{D}^2-4\mub\big)\bar{\phi}_{\ad(n)}=0~,  \label{OS3.b}
\end{align} 
\end{subequations}
respectively, whilst \eqref{OS1} remains the same. In the case when  $m=n=0$, the on-shell superfields are discussed in sections
\ref{WZ section} and \ref{section5.5}.

In analogy with \eqref{Mphys}, we define the physical mass $M_{\text{phys}}$ through
\begin{align}
M^2_{\text{phys}}= M^2-\lambda_{(1,m,n)}\mu\mub~. \label{SMphys}
\end{align}
As discussed in section \ref{section review}, the pseudo-mass of an on-shell non-supersymmetric field is subject to a unitarity bound. Since the on-shell supermultiplet $\phi_{\a(m)\ad(n)}$ contains a multitude of such fields, this in turn induces a unitarity bound on the pseudo-mass of $\phi_{\a(m)\ad(n)}$:
\begin{align}
M^2_{\text{phys}}\geq 0  \quad \implies \quad M^2\geq \lambda_{(1,m,n)}\mu\mub~. \label{Sbound}
\end{align}

By going to components, we will see in section \ref{section 3.5} that the on-shell supermultiplet $\phi_{\a(m)\ad(n)}$ furnishes the representation $\mathfrak{S}\big(E_0(M),\frac{1}{2}(m+n+1)\big)$ of $\sOSp (1|4)$, where
\begin{align}
E_0 (M) =1  + \frac{1}{2}
\sqrt{ 4\frac{M^2}{\mu\mub}-(m+n)(m+n+4) +1} ~. \label{Energy}
\end{align}


\subsection{Massless supermultiplets}\label{section 3.1}

Massless supermultiplets correspond to those on-shell superfields 
carrying pseudo-mass
\begin{align}
M^2=\lambda_{(1,m,n)}\mu\mub~. \label{massless}
\end{align} 
There are at least two different angles from which this can be understood. 

The first is the fact that the system of equations \eqref{OS} with 
$m\geq n >0$ and with 
$M^2$ satisfying \eqref{massless} is compatible with  gauge transformations of the form \eqref{GT},
\begin{subequations} \label{MasslessGI}
\begin{align}
\delta_{\zeta,\xi}\phi_{\a(m)\ad(n)}=\bar{\mathcal{D}}_{(\ad_1}
\zeta_{\a(m)\ad_2\dots\ad_n)}+\mathcal{D}_{(\a_1}\xi_{\a_2\dots\a_m)\ad(n)}~,\label{MasslessGIa}
\end{align} 
where the gauge parameters $\zeta_{\a(m)\ad (n-1)}$ and 
$\x_{\a(m-1) \ad (n)}$ are TLAL and obey the constraints
\bea 
\cD_{(\a_1}{}^\bd \x_{\a_2 \dots \a_m) \bd \ad (n-1)} &=&\phantom{-} \ri (n+1) \m \zeta_{\a(m)\ad (n-1)}~, \label{GoodDaySirA}\\
\cD^\b{}_{(\ad_1} \z_{\b \a(m-1)  \ad_2 \dots  \ad_n)} 
&=& -\ri (m+1) \bar \m \x_{\a(m-1)\ad (n)}~.\label{GoodDaySirB}
\eea
\end{subequations}
These on-shell conditions imply that $\zeta_{\a(m)\ad (n-1)}$ and 
$\x_{\a(m-1) \ad (n)}$ satisfy the equations  
\bea
 \big(\mathbb{Q}-\lambda_{(1,m,n)}\mu\mub\big)\z_{\a(m)\ad(n-1)}=0~, \qquad
  \big(\mathbb{Q}-\lambda_{(1,m,n)}\mu\mub\big)\x_{\a(m-1)\ad(n)}=0~.\label{Blah3}
\eea
Thus, from \eqref{Blah3} and the gauge variation of \eqref{OS1}, we see that the gauge parameters are only non-zero if $M$ satisfies \eqref{massless}. 

When $m=n=s$ one can consistently impose the reality condition $H_{\a(s)\ad(s)}:=\phi_{\a(s)\ad(s)}=\bar{H}_{\a(s)\ad(s)}$, whereupon the gauge transformations \eqref{MasslessGIa} take the form
\begin{subequations} \label{massless4.9}
\begin{align}
\delta_{\zeta}H_{\a(s)\ad(s)}=\bar{\mathcal{D}}_{(\ad_1}
\zeta_{\a(s)\ad_2\dots\ad_s)}-\mathcal{D}_{(\a_1}\bar \z_{\a_2\dots\a_s)\ad(s)}~,\label{MasslessGIaa}
\end{align} 
whilst \eqref{GoodDaySirA} and \eqref{GoodDaySirB} become
\begin{align}
\cD^\b{}_{(\ad_1} \z_{\b\a(s-1)\ad_2 \dots \ad_s)} &= \phantom{-}\ri (s+1) \mub \bar{\zeta}_{\a(s-1)\ad (s)}~,\label{Blah1}\\
\cD_{(\a_1}{}^\bd \bar{\z}_{\a_2 \dots \a_s) \bd \ad (s-1)} &= -\ri (s+1) \m \zeta_{\a(s)\ad (s-1)}~. \label{Blah2}
\end{align}
\end{subequations}

The presence of the gauge freedom \eqref{MasslessGI} means that the 
 $\sOSp (1|4)$ representation  
\eqref{OS} with $M^2$ given by \eqref{massless} is not irreducible. To obtain an irreducible representation,  the space of TLAL superfields 
\eqref{OS} and \eqref{massless} has to be factorised
with respect to the gauge modes. More specifically, two superfields $\phi_{\a(m)\ad(n)}$ and $\widetilde \phi_{\a(m)\ad(n)}$ are said to be equivalent if they differ by a gauge transformation \eqref{MasslessGI}. The genuine massless representation is realised on the quotient space of
the space of on-shell superfields 
\eqref{OS} and \eqref{massless} with respect to this equivalence relation. 
The gauge degrees of freedom are automatically eliminated if one works with 
the gauge-invariant chiral field strength 
 \eqref{HSW1}, 
\bea
\mathfrak{W}_{\a(m+n+1)}(\f)&= -\frac{1}{4}\big(\bar{\mathcal{D}}^2-4\mu\big)\mathcal{D}_{(\a_1}{}^{\bd_1}\cdots\mathcal{D}_{\a_n}{}^{\bd_n}\mathcal{D}_{\a_{n+1}}\f_{\a_{n+2}\dots\a_{m+n+1})\bd(n)} ~,
\label{CFS4.8}
\eea
instead of the prepotential $\phi_{\a(m)\ad(n)}$. 
 Making use of the constraint \eqref{OS2} satisfied by 
 $\f_{\a(m)\ad(n)}$, one may show that 
 $\mathfrak{W}_{\a(m+n+1)}(\f)$ is constrained by
\begin{align}
0=\mathcal{D}^{\b}\mathfrak{W}_{\b \a(m+n)}(\f)\quad\implies\quad 0=\big(\mathbb{Q}-\lambda_{(1,m,n)}\mu\mub\big)\mathfrak{W}_{\a(m+n+1)}(\f)~,
\label{4.9}
\end{align}
where the second relation follows from  the chirality of the field strength. Once again, the latter is consistent with \eqref{OS1} only if $M$ satisfies \eqref{massless}.
The description in terms of $\mathfrak{W}_{\a(m+n+1)}(\f)$ provides the second way to formulate massless dynamics. 

Let us return to the gauge-invariant chiral field strengths \eqref{3.16} and \eqref{3.18}, which 
originate in the massless models for the superspin-$(s+\hf)$ and superspin-$s$ multiplets in AdS$_4$ \cite{KS94}, respectively. As was demonstrated in \cite{KS94}, on the mass shell they satisfy the equations 
\begin{subequations}
\bea
\mathcal{D}^{\b}\mathfrak{W}_{\b \a(2s)}(H) &=&0~, \\
\qquad 
\mathcal{D}^{\b}\mathfrak{W}_{\b \a(2s-1)}(\J) &=&0~,
\eea
\end{subequations}
which are exactly the on-shell constraints \eqref{4.9}.


\subsection{Partially massless supermultiplets} \label{section 3.2}

The transverse projection operators for non-supersymmetric tensor fields in AdS$_4$ were constructed recently in \cite{KP20}. There it was observed that the poles of the projectors were intimately related to (partially) massless fields. The structure of on-shell $\mathcal{N}=1$ supermultiplets containing partially massless fields
was recently discussed 
 at the component level in \cite{G-SHR, BKSZ}.
 However, their realisations in terms of on-shell superfields are not yet known. In this section we address this gap.     

In the spirit of \cite{KP20}, it is natural to expect that the partially-massless supermultiplets are associated with the poles of the TLAL projectors \eqref{projectors}. This motivates the definition of a partially massless superfield to be one satisfying the on-shell constraints \eqref{OS} such that the pseudo-mass takes one of the values \eqref{SPM},
\begin{align}
M^2=\lambda_{(t,m,n)}\mu\mub~,\qquad 1\leq t \leq \text{min}(m+1,n+1)~. \label{MS}
\end{align}
We will say that $\phi_{\a(m)\ad(n)}$ carries `super-depth' $t$. In section \ref{section 3.5} we show that, with the above definitions, the surviving component fields of an on-shell partially massless superfield with super-depth $t>1$ are themselves partially massless with depths $t$ or $t-1$.  We would like to point out that $\lambda_{(t,m,n)}$ with $1\leq t \leq \text{max}(m+1,n+1)$ appear in the poles of the projectors \eqref{projectors}.  The reason that we have chosen $t=\text{min}(m+1,n+1)$ as an upper bound on $t$ in the definition \eqref{MS} is because, as we will see, there is no gauge symmetry present for $\text{min}(m+1,n+1) < t \leq \text{max}(m+1,n+1)$.

From \eqref{massless}, we see that partially massless supermultiplets with the minimal super-depth of $t=1$ correspond to massless representations and are hence unitary. However, partially massless supermultiplets whose super-depth lies within the range $2\leq t \leq \text{min}(m+1,n+1)$ have negative physical mass and describe non-unitary representations of $\sOSp(1|4)$. This may be seen from the inequality
\begin{align}
\lambda_{(1,m,n)} > \lambda_{(t,m,n)}~ , \qquad 2\leq t \leq \text{min}(m+1,n+1)~.
\label{PMbound}
\end{align}
 To distinguish the latter we may sometimes refer to them as true partially massless supermultiplets. In analogy with \eqref{NSPMrep}, we will denote 
 the true partially massless representation with super-depth $t$ and Lorentz type $(m/2,n/2)$
 by 
 \begin{align}
 \mathfrak{P}(t,m,n)~, \qquad 2\leq t \leq \text{min}(m+1,n+1)~.
 \end{align}
According to \eqref{Energy}, such a representation carries minimal energy $E_0=\frac{1}{2}(m+n+1)-t+2$.

In the previous section it was observed that on-shell, \eqref{massless} is the only pseudo-mass value compatible with the massless gauge symmetry \eqref{MasslessGIa}. For true partially massless supermultiplets the story is considerably more complicated. This is because the correct gauge symmetry at the superspace level is not yet known, much less the gauge invariant actions and field strengths. 
However, by making use of the superprojectors proposed in section \ref{main}, it is possible to systematically derive the most general gauge symmetry compatible with the on-shell conditions \eqref{OS}. In appendix B this procedure is carried out in detail for the real supermultiplet $H_{\a\ad}$. The results of this analysis, and that of the supermultiplet $H_{\a(2)\ad(2)}$ obtained by analogy, are summarised below. 

\begin{table}[h]
\begin{center}
\begin{tabular}{|c|c|c|c|}
\hline
~Super-depth~  & $t=1$     & $t=2$ & $t=3$  \Tstrut\Bstrut\\ \hline
  $\delta H_{\a\ad}$        &  $\bar{\mathcal{D}}_{\ad}\z_{\a}-\mathcal{D}_{\a}\bar{\z}_{\ad}$   & $\mathcal{D}_{\a\ad}(\s+\bar{\s})$  & -- \Tstrut\\
  \hline
  $\delta H_{\a(2)\ad(2)}$    & $\bar{\mathcal{D}}_{\ad}\z_{\a(2)\ad}-\mathcal{D}_{\a}\bar{\z}_{\a\ad(2)}$   & $\mathcal{D}_{\a\ad}\big(\bar{\mathcal{D}}_{\ad}\xi_{\a}-\mathcal{D}_{\a}\bar{\xi}_{\ad}\big)$     & $\mathcal{D}_{\a\ad}\mathcal{D}_{\a\ad}(\eta+\bar{\eta})$\\ \hline
\end{tabular}
\end{center} 
\vspace{-15pt} 
\caption{Gauge symmetry for lower-rank PM supermultiplets.}
\label{table 3}
\end{table}

In table \ref{table 3},\footnote{In table \ref{table 3}, and in the remainder of this subsection, we employ the notational convention \eqref{symmnot}. } the most general gauge symmetries for the aforementioned supermultiplets are given. In both cases it turns out that a gauge symmetry is present only when the pseudo-mass takes the values specified in \eqref{MS}. The gauge parameters are on-shell in the sense that they each possess the same pseudo-mass as its parent gauge field and that they satisfy certain irreducibility constraints. Specifically, in the super-depth $t=1$ case, the gauge parameters $\z_\a$ and $\z_{\a(2) \ad} $ in table \ref{table 3} are LTAL and TLAL, respectively,  and obey the reality conditions \eqref{Blah1} and \eqref{Blah2}. The gauge parameter $\x_\a$, which corresponds to $H_{\a(2)\ad(2)}$ with super-depth $t=2$, is LTAL and obeys the reality condition 
\bea
\mathcal{D}_{\ad}{}^{\b}\xi_{\b}=3\text{i}\mub\bar{\xi}_{\ad}~.
\eea
In the case of maximal super-depth, 
the gauge parameters $\s$ and $\eta$ are chiral and  obey 
the  reality conditions
\begin{subequations} \label{suppGb}
\bea
-\frac{1}{4}\big(\mathcal{D}^2-4\mub\big)\s+2\mub\bar{\s}&=&0~, \\
 -\frac{1}{4}\big(\mathcal{D}^2-4\mub\big)\eta+3\mub\bar{\eta}&=&0~.
\eea
\end{subequations}
These conditions are the equations of motion which follow from various Wess-Zumino models (c.f. \eqref{EoMWZ}).

The method used to derive the above results (see appendix \ref{B.2} for the details) is quite general in that it deduces all types of gauge symmetry an on-shell supermultiplet can possess, and at which mass values they appear. However, for higher-rank multiplets it quickly becomes infeasible due to the computational expense. Nevertheless, from the results of the above analysis, we are now in a position to make an ansatz for the gauge transformations of the  half-integer superspin-$(s+\frac{1}{2})$ multiplets  $H^{(t)}_{\a(s)\ad(s)}$. For true partially-massless multiplets with super-depth $2\leq t \leq s+1$, they take the form\footnote{Interestingly, the higher-depth gauge transformations \eqref{Blah7} are different to those proposed for the generalised superconformal multiplets in \cite{KPR2}. The latter transformations possess the same functional form as the second group of terms in \eqref{decomp2}.
}
\begin{subequations}
\begin{align}
2\leq t \leq s~&: ~\qquad \delta_{\xi}H^{(t)}_{\a(s)\ad(s)}=\big(\mathcal{D}_{\a\ad}\big)^{t-1}\big(\bar{\mathcal{D}}_{\ad}\xi_{\a(s-t+1)\ad(s-t)}-\mathcal{D}_{\a}\bar{\xi}_{\a(s-t)\ad(s-t+1)}\big)~,\label{Blah7}\\[5pt]
t=s+1~&:~\qquad  \delta_{\s}H^{(s+1)}_{\a(s)\ad(s)}=\big(\mathcal{D}_{\a\ad}\big)^{s}\big(\s+\bar{\s}\big)~. \label{Blah8}
\end{align}
\end{subequations}
The system of equations \eqref{OS} and \eqref{MS} is invariant under the transformations \eqref{Blah7} as long as the gauge parameters are TLAL and satisfy the reality conditions
\begin{subequations}\label{Blah5}
\begin{align}
\cD^\b{}_{\ad} \xi_{\b\a(s-t)\ad(s-t)} &= \phantom{-}\ri (s+1) \mub \bar{\xi}_{\a(s-t)\ad (s-t+1)}~, \\
\cD_{\a}{}^{\bd} \bar{\xi}_{\a(s-t)\ad(s-t)\bd} &= -\ri (s+1) \mu \xi_{\a(s-t+1)\ad (s-t)}~. 
\end{align}
\end{subequations}
The same is true for the transformations \eqref{Blah8} given that $\s$ is chiral, $\bar{\mathcal{D}}_{\ad}\s=0$, and that it satisfies the equations
\begin{subequations}\label{BLAH7}
\begin{align}
-\frac{1}{4}\big(\mathcal{D}^2-4\mub\big)\s+(s+1)\mub\bar{\s}&=0~,\\
-\frac{1}{4}\big(\bar{\mathcal{D}}^2-4\mu\big)\bar{\s}+(s+1)\mu\s &=0~.
\end{align}
\end{subequations}
It may be shown that \eqref{Blah5} and \eqref{BLAH7} imply that each gauge parameter satisfies the same mass-shell equation as its corresponding gauge field, as required. We note that upon substituting $t=1$ into \eqref{Blah7} and \eqref{Blah5}, one recovers the massless gauge transformations \eqref{massless4.9} with $\xi_{\a(s)\ad(s-1)}\equiv \z_{\a(s)\ad(s-1)}$.



\subsection{Massive supermultiplets} \label{section 3.3}

We define a massive superfield $\phi_{\a(m)\ad(n)}$ to be one satisfying the on-shell conditions \eqref{OS} and whose pseudo-mass satisfies
\begin{align}
M^2 > \lambda_{(1,m,n)}\mu\mub~,  \label{MSC}
\end{align}
but is otherwise arbitrary. By virtue of \eqref{Sbound} this definition ensures that the physical mass is positive and that the corresponding representation of $\sOSp(1|4)$ is unitary. 


\subsection{Equivalent representations and reality conditions}

Given two positive integers $m$ and $n$, with $m+n \equiv 2s$, 
we denote by  $\cL_{(m,n)}^{[M]} $ the space of on-shell superfields  $\phi_{\a(m)\ad(n)}$ 
satisfying the conditions \eqref{OS}. The following proposition holds: 
Provided the pseudo-mass satisfies
\begin{align}
M^2 \neq \l_{(t, m,n)}~,\qquad 1\leq t \leq \text{max}(m+1,n+1)~, \label{equiv55}
\end{align}
the $\sOSp(1|4)$ representations on the functional spaces  $\cL_{(2s,0)}^{[M]},~ 
\cL_{(2s-1,1)}^{[M]} , \dots 
 \cL_{(1,2s-1)}^{[M]}, ~  \cL_{(0, 2s)}^{[M]} $ are equivalent to 
 $\mathfrak{S} \big( E_0 (M) , s +\hf \big)$, where $E_0(M)$ is given by \eqref{Energy}.

To prove the above claim, we consider  an arbitrary superfield $\phi_{\a(m)\ad(n)} \in \cL_{(m,n)}^{[M]}$ and  associate with it
 the following descendants 
\begin{subequations}\label{4.16}
\bea
 \phi_{\a(m)\ad(n)} ~& \to &~  \j_{\a(m+1)\ad(n-1)}
 := \cD_{\a_{m+1} }{}^\bd  \phi_{\a(m)\bd \ad(n-1)} 
\label{4.16a}
~, \\
 \phi_{\a(m)\ad(n)} ~& \to &~  \c_{\a(m-1)\ad(n+1)}
 := \cD^{\b }{}_{\ad_{n+1}}  \phi_{\b\a(m-1)\ad(n)} ~.
 \label{4.16b}
 \eea
 \end{subequations} 
 It is obvious that $ \j_{\a(m+1)\ad(n-1)}$ is completely symmetric in its undotted indices, 
 while  $\c_{\a(m-1)\ad(n+1)} $ is completely symmetric in its dotted indices. 
 The descendant $ \j_{\a(m+1)\ad(n-1)}$ proves to obey the conditions \eqref{OS}  
 if $n\neq 1$. In the $n=1$ case,  $ \j_{\a(m+1)\ad(n-1)}$ is an on-shell superfield of the type \eqref{OS3.a}. The descendant $ \c_{\a(m-1)\ad(n+1)}$ has analogous properties. 
 Therefore the relations \eqref{4.16a} and \eqref{4.16b} define linear mappings
from  $\cL_{(m,n)}^{[M]} $ to 
 $\cL_{(m+1,n-1)}^{[M]} $ and  $\cL_{(m-1,n+1)}^{[M]} $, respectively. 
 
Making use of the identities \eqref{A.4}
leads to the relations
\begin{subequations}\label{4.17}
\bea
 \cD^\b{}_\gd \cD_{\b }{}^\bd  \phi_{\a(m)\bd \ad(n-1)} &=& 
 \big(M^2 - \l_{(n+1, m, n)} \m \bar \m  \big)\phi_{\a(m)\gd \ad(n-1)} ~,\\
  \cD_\g{}^\bd \cD^{\b }{}_\bd  \phi_{\b\a(m-1) \ad(n)} &=& 
 \big(M^2 - \l_{(m+1, m, n)} \m \bar \m  \big)\phi_{\g \a(m-1) \ad(n)}~.
\eea
\end{subequations}
In conjunction with the identities
\bea
\l_{(n+1, m, n)} &=& \l_{(m+2,m+1,n-1)}~, \qquad
 \l_{(m+1, m, n)} = \l_{(n+2,m-1,n+1)}~,
\eea
the relations \eqref{4.17} tell us  
 that the linear transformations \eqref{4.16a} and \eqref{4.16b} 
are one-to-one and onto, as long as 
$M^2 \neq \l_{(n+1, m, n)} \m \bar \m$ and $M^2 \neq  \l_{(m+1, m, n)} \m \bar \m$.

Let us introduce the linear maps
\bea
\D_{(m,n)} : \cL_{(m,n)}^{[M]}  \to 
 \cL_{(m+1,n-1)}^{[M]} ~, \qquad 
\widetilde  \D_{(m,n)} : \cL_{(m,n)}^{[M]}  \to 
 \cL_{(m-1,n+1)}^{[M]} ~,
\eea
defined by
\begin{subequations}
\bea
 \phi_{\a(m)\ad(n)} ~& \to &~  \f_{\a(m+1)\ad(n-1)}
 :=  \big(\D_{(m,n)} \big)_{\a_{m+1} }{}^\bd  \phi_{\a(m)\bd \ad(n-1)} ~, \\
  \phi_{\a(m)\ad(n)} ~& \to &~  \f_{\a(m-1)\ad(n+1)}
 :=  \big(\widetilde \D_{(m,n)} \big)^\b{}_{\ad_{n+1} }  \phi_{\b\a(m-1)\bd \ad(n)} ~, 
\eea
\end{subequations}
where we have introduced the dimensionless operators
\bea
 \big(\D_{(m,n)} \big)_{\a\ad} = \frac{\mathcal{D}_{\a\ad}}{\sqrt{\mathbb{Q}-\lambda_{(n+1,m,n)}\mu\mub}}~, \qquad 
 \big(\widetilde{\D}_{(m,n)} \big)_{\a\ad} = \frac{\mathcal{D}_{\a\ad}}{\sqrt{\mathbb{Q}-\lambda_{(m+1,m,n)}\mu\mub}}~.~~~ \label{indexconverters}
 \eea
The above relations are equivalent to 
\bea
\widetilde \D_{(m+1,n-1)} \D_{(m,n)} \Big|_{\cL_{(m,n)}^{[M]}  } = {\mathbbm 1}~,\qquad 
 \D_{(m-1,n+1)} \widetilde \D_{(m,n)} \Big|_{\cL_{(m,n)}^{[M]}  } = {\mathbbm 1}~.
\eea
Therefore, all the superfields 
\bea
 \phi_{\a(2s)}, ~ \phi_{\a(2s-1)\ad} ,~ \dots   , ~\phi_{\a(m)\ad(n)} ,~ \dots , 
~  \phi_{\a\ad(2s-1)}, ~ \phi_{\ad(2s)} 
 \eea
also satisfy \eqref{OS} and hence furnish the representation $\mathfrak{S} \big( E_0 (M) , s +\hf \big)$.
 
 In proving the above assertion it was assumed that the pseudo-mass of $\phi_{\a(m)\ad(n)}$
 satisfies \eqref{equiv55}. This was to ensure that when $\phi_{\a(m)\ad(n)}$ is on the mass-shell, the operators \eqref{indexconverters} are well defined. When $\phi_{\a(m)\ad(n)}$ is partially-massless with super-depth $t$, at some stage the maps between tensor types become ill defined. Nevertheless, in this case it may be shown that the corresponding representation may be equivalently realised on any of the functional spaces $\mathcal{L}^{[M]}_{(m+n-t+1,t-1)},\mathcal{L}^{[M]}_{(m+n-t,t)},\dots,\mathcal{L}^{[M]}_{(t,m+n-t)},\mathcal{L}^{[M]}_{(t-1,m+n-t+1)}$.
 
 In general, $ \phi_{\a(m)\ad(n)} $ and its conjugate $\bar  \phi_{\a(n)\ad(m)} $ 
 describe two equivalent representations of  $\sOSp(1|4)$. In order to obtain a single 
 representation, we need to impose a reality condition on $ \phi_{\a(m)\ad(n)} $. 
 In the case when $m+n \equiv 2s $ is even,  $ \phi_{\a(m)\ad(n)} $ 
 can represented  by  $ \phi_{\a(s)\ad(s)} $ which we require to be real,
 \bea
\bar  \phi_{\a(s)\ad(s)}  =\phi_{\a(s)\ad(s)}~.
\eea
If $m+n \equiv 2s-1 $ is odd,  $ \phi_{\a(m)\ad(n)} $ can be represented 
by $\phi_{\a(s)\ad(s-1)}$. In this case a reality condition can be chosen in the form of a
Dirac-type pair of equations
 \begin{subequations} \label{Dirac}
\begin{align}
\mathcal{D}_{(\ad_1}{}^{\b}\phi_{\b\a(s-1)\ad_2\dots\ad_s)}&=\re^{\ri \g} 
M_{(s+1)}\bar{\phi}_{\a(s-1)\ad(s)}~,\\
 \mathcal{D}_{(\a_1}{}^{\bd}\bar{\phi}_{\a_2\dots\a_s)\ad(s-1)\bd}&=
 \re^{-\ri \g} M_{(s+1)}\phi_{\a(s)\ad(s-1)}~,
\end{align}
\end{subequations}
where $M^2_{(s+1)}:=M^2-\lambda_{(s+1,s,s-1)}\mu\mub$ and $\g$ is a constant real phase. 
The pair of equations \eqref{Dirac} lead to the same mass-shell equation \eqref{OS1}.   
 
 The above consideration offers a simple way to re-derive the superprojectors 
 \eqref{projectors}. Consider an off-shell superfield $\F_{\a(m) \ad (n)}$ and 
 convert its dotted indices into undotted 
 \bea
\F_{\a(m) \ad (n)} ~\to ~ \F_{\a(m+n)} :=  \big(\D_{(m+n-1,1)} \big)_{(\a_1}{}^{\bd_1} \dots  
 \big(\D_{(m,n)} \big)_{\a_n}{}^{\bd_n} \F_{\a_{n+1} \dots \a_{n+m} )\bd(n)} ~.
 \eea
 Next we apply the superprojector $\Pi^{(m+n)}$, eq. \eqref{3.33a}, to $\F_{\a(m+n)}$.
 The resulting superfield $\vf_{\a(m+n)}:=   \Pi^{(m+n)}  \F_{\a(m+n)}$ has the properties listed in 
 \eqref{3.34}. Finally we convert $n$ undotted indices of $ \vf_{\a(m+n)}$ back into dotted ones, 
 \bea
 \vf_{\a(m+n)}~ \to~ \vf_{\a(m) \ad(n) } 
: =  \big(\widetilde \D_{(m+n,0)} \big)^{\b_1}{}_{\ad_1 } \dots  
 \big(\widetilde \D_{(m+1,n-1)} \big)^{\b_n}{}_{\ad_{n} } \vf_{\b (n) \a(m) }~.
\eea
One observes that $\vf_{\a(m) \ad(n) } $ coincides with $\Pi^{(m,n)}\Phi_{\a(m)\ad(n)}$,
eq. \eqref{proj1}.


\section{Component analysis} \label{section 3.5}

We now turn to the component analysis of the on-shell superfield $\phi_{\a(m)\ad(n)}$ satisfying the conditions  \eqref{OS} with arbitrary $M^2$. 
For this we make use of the bar-projection of a tensor superfield ${\mathfrak V} ={\mathfrak V}(x,\q, \bar \q)$ (with suppressed indices)
which is defined as usual:
\bea
{\mathfrak V}|:= {\mathfrak V}(x,\q, \bar \q )\big|_{\q =\bar \q=0}~.
\eea 
The supermultiplet of fields associated with $\mathfrak V$ is defined to consist 
of all independent fields contained in 
$\big\{ {\mathfrak V}|, \cD_{ \a} {\mathfrak V}|, \bar \cD_{ \ad} {\mathfrak V}|, \dots \big\}$.
The covariant derivative of AdS$_4$, $\nabla_{a}$,  
is related to $\cD_{a}$ according to the rule
\begin{align}
\nabla_{a} \cV  :=(\mathcal{D}_{a} {\mathfrak V} )|~, \qquad \cV := {\mathfrak V}|~.
\end{align}
In practice, we assume a Wess-Zumino gauge condition to be imposed on the geometric objects in \eqref{33.1}
such that the background geometry 
is purely bosonic, 
\bea
\cD_a| :=  E_a{}^M |\partial_M  + \frac{1}{2}\,\O_a{}^{bc} | M_{bc}
=e_a{}^m \partial_m  + \frac{1}{2}\,\o_a{}^{bc} M_{bc} = \nabla_a~.
\eea
The supersymmetry transformation of the component fields of $\mathfrak V$ 
is computed according to the rule \eqref{A.14}

Below we make use of the non-supersymmetric AdS quadratic Casimir operator 
\eqref{Cas2}
which is related to the Casimir \eqref{Cas1} via $\mathcal{Q} \cV =\Big[ \Big( \mathbb{Q}-\frac{1}{4}\big(\mu\mathcal{D}^2+\mub\bar{\mathcal{D}}^2\big) \Big) {\mathfrak V}\Big]\Big|$.


\subsection{Massive supermultiplets}

 In general there are four non-vanishing independent complex component fields,
\begin{subequations}
\begin{align}
A_{\a(m)\ad(n)}&:=\phi_{\a(m)\ad(n)}|~,\\
B_{\a(m+1)\ad(n)}&:=\mathcal{D}_{(\a_1}\phi_{\a_2\dots\a_{m+1})\ad(n)}|~,\\
C_{\a(m)\ad(n+1)}&:=\bar{\mathcal{D}}_{(\ad_1}\phi_{\a(m)\ad_2\dots\ad_{n+1})}|~,\\
E_{\a(m+1)\ad(n+1)}&:=\bigg(\frac{1}{2}\big[\mathcal{D}_{(\a_1},\bar{\mathcal{D}}_{(\ad_1}\big]-\text{i}\frac{m-n}{m+n+2}\mathcal{D}_{(\a_1(\ad_1}\bigg)\phi_{\a_2\dots\a_{m+1})\ad_2\dots\ad_{n+1})}|~.
\end{align}
\end{subequations}
They have each been defined in such a way that they are transverse 
\begin{subequations}
\begin{align}
0&=\nabla^{\b\bd}A_{\b\a(m-1)\bd\ad(n-1)}~,\label{PMA1}\\
0&=\nabla^{\b\bd}B_{\b\a(m)\bd\ad(n-1)}~,\label{PMB1}\\
0&=\nabla^{\b\bd}C_{\b\a(m-1)\bd\ad(n)}~,\label{PMC1}\\
0&=\nabla^{\b\bd}E_{\b\a(m)\bd\ad(n)}~,\label{PMD1}
\end{align}
\end{subequations}
which may be shown by making use of the on-shell conditions \eqref{OS}. Furthermore, one may show that the bottom and top components satisfy the mass-shell equations
\begin{subequations} \label{MS1}
\begin{align}
0&=\Big(\mathcal{Q}-\big[M^2-\frac{1}{2}(m+n+4)\mu\bar{\mu}\big]\Big)A_{\a(m)\ad(n)}~,\\
0&=\Big(\mathcal{Q}-\big[M^2+\frac{1}{2}(m+n)\mu\bar{\mu}\big]\Big)E_{\a(m+1)\ad(n+1)}~,
\end{align}
\end{subequations}
 whilst the two middle component fields satisfy the differential equations
\begin{subequations}
\bea
0&=&\phantom{-}\ri \m \nabla_{(\ad_1}{}^{\b}B_{\b\a(m)\ad_2\dots\ad_{n+1})}
+
\Big(\mathcal{Q}-\big[M^2-\frac{1}{2}(m-n+3)\mu\bar{\mu}\big]\Big)C_{\a(m)\ad(n+1)}
~, \label{EOM1}\\
0&=& -\ri \mub \nabla_{(\a_1}{}^{\bd}C_{\a_2\dots\a_{m+1})\ad(n)\bd}
+
\Big(\mathcal{Q}-\big[M^2-\frac{1}{2}(n-m+3)\mu\bar{\mu}\big]\Big)B_{\a(m+1)\ad(n)}~. 
~~~\label{EOM2}
\eea
\end{subequations}
The latter imply
 that only one of the two fields $B_{\a(m+1)\ad(n)}$ and $C_{\a(m)\ad(n+1)}$ is independent, and lead to
the higher derivative mass-shell equations
\begin{subequations}\label{MS2}
\begin{align}
0&=\Big(\mathcal{Q}-M^2_{+}\Big)\Big(\mathcal{Q}-M^2_{-}\Big)B_{\a(m+1)\ad(n)}~,\\
0&=\Big(\mathcal{Q}-M^2_{+}\Big)\Big(\mathcal{Q}-M^2_{-}\Big)C_{\a(m)\ad(n+1)}~,
\end{align}
\end{subequations}
where we have denoted
\begin{align}
M^2_{\pm}&= M^2-\mu\bar{\mu}\pm\frac{\mu\mub}{2}\sqrt{ 4\frac{M^2}{\mu\mub}-(m+n)(m+n+4)+1}~.
\end{align}

Using the relation \eqref{Energy-Mass}, one may confirm that the above component results 
are in agreement with the decomposition \eqref{E1},  
\begin{align}
\mathfrak{S}\big({E}_0,s\big) =
D\big(E_0+ \frac{1}{2}, s- \frac{1}{2}\big)  &\oplus D\big( E_0, s\big) \oplus D \big(E_0+1, s\big) \oplus  D\big(E_0+ \frac{1}{2}, s+ \frac{1}{2}\big)~, \label{Blah4}
\end{align}
as dictated by representation theory. 
Here 
$s=\frac{1}{2}(m+n+1)$ and $E_0\equiv E_0(M)$ is defined according to $\eqref{Energy}$. 


\subsection{Partially massless supermultiplets} \label{section 3.6}
In this section we restrict our attention to on-shell true partially massless supermultiplets with super-depth $2\leq t \leq \text{min}(m+1,n+1)$. The massless case with $t=1$ will be considered separately in the next section. Upon specifying the pseudo-mass to be given by \eqref{MS}, one can show that the mass-shell equations \eqref{MS1} and \eqref{MS2} become
\begin{subequations}
\begin{align}
0&=\Big(\mathcal{Q}-\tau_{(t-1,m,n)}\mu\bar{\mu}\Big)A_{\a(m)\ad(n)}~,\label{PMA2}\\
0&=\Big(\mathcal{Q}-\tau_{(t-1,m+1,n)}\mu\bar{\mu}\Big)\Big(\mathcal{Q}-\tau_{(t,m+1,n)}\mu\bar{\mu}\Big)B_{\a(m+1)\ad(n)}~,\label{PMB2}\\
0&=\Big(\mathcal{Q}-\tau_{(t-1,m,n+1)}\mu\bar{\mu}\Big)\Big(\mathcal{Q}-\tau_{(t,m,n+1)}\mu\bar{\mu}\Big)C_{\a(m)\ad(n+1)}~,\label{PMC2}\\
0&=\Big(\mathcal{Q}-\tau_{(t,m+1,n+1)}\mu\bar{\mu}\Big)E_{\a(m+1)\ad(n+1)}~. \label{PMD2}
\end{align}
\end{subequations}
Here $\tau_{(t,m,n)}$ are the non-supersymmetric partially massless values \eqref{PMval}, which are related to the supersymmetric ones \eqref{SPM} through
\begin{align}
\lambda_{(t,m,n)}=\tau_{(t-1,m,n)}+\frac{1}{2}(m+n+4)~.
\end{align}

In accordance with the discussion of on-shell partially massless fields given in section \ref{section NSPM}, the following remarks hold:
\begin{itemize}
\item The pair of equations \eqref{PMA1} and \eqref{PMA2} admits a depth $t_{A}=t-1$ gauge symmetry
\begin{align}
\delta_{\zeta}A_{\a(m)\ad(n)}=\nabla_{(\a_1(\ad_1}\cdots\nabla_{\a_{t_A}\ad_{t_A}}\zeta_{\a_{t_A+1}\dots\a_{m})\ad_{t_A+1}\dots\ad_{n})}
\end{align}
when $2\leq t \leq \text{min}(m,n)$.
\item The pair of equations \eqref{PMD1} and \eqref{PMD2} admits a depth $t_{E}=t$ gauge symmetry
\begin{align}
\delta_{\xi}E_{\a(m+1)\ad(n+1)}=\nabla_{(\a_1(\ad_1}\cdots\nabla_{\a_{t_E}\ad_{t_E}}\xi_{\a_{t_E+1}\dots\a_{m+1})\ad_{t_E+1}\dots\ad_{n+1})}
\end{align}
when $2\leq t \leq \text{min}(m+1,n+1)$.

\item Equation \eqref{PMB1} and the first branch of \eqref{PMB2} admit a depth $t_B=t-1$ 
\begin{align}
\delta_{\rho}B_{\a(m+1)\ad(n)}=\nabla_{(\a_1(\ad_1}\cdots\nabla_{\a_{t_B}\ad_{t_B}}\rho_{\a_{t_B+1}\dots\a_{m+1})\ad_{t_B+1}\dots\ad_{n})}\label{GSB2}
\end{align}
gauge symmetry when $2\leq t\leq \text{min}(m+1,n+1)$. 

\item Equation \eqref{PMB1} and the second branch of \eqref{PMB2} admit a depth $t'_B=t$ 
\begin{align}
\delta_{\rho}B_{\a(m+1)\ad(n)}=\nabla_{(\a_1(\ad_1}\cdots\nabla_{\a_{t'_B}\ad_{t'_B}}\rho_{\a_{t'_B+1}\dots\a_{m+1})\ad_{t'_B+1}\dots\ad_{n})}\label{GSB1}
\end{align}
gauge symmetry when $2\leq t \leq \text{min}(m+1,n)$.

\item Comments similar to those above regarding $B_{\a(m+1)\ad(n)}$, apply to $C_{\a(m)\ad(n+1)}$.
\item The above gauge symmetries hold only for gauge parameters which are transverse and which satisfy the same mass-shell equation as the parent gauge field. 
\end{itemize}

One can reverse the logic and ask the question: for what values of $M^2$ does there appear a gauge symmetry (of any depth) at the component level? The answer is precisely those values which appear in the superprojectors \eqref{SPM} (though, as above, special care must be taken when deducing the upper and lower bounds on the range of possible depths). 

From the above component analysis one can see that the true partially massless representation $\mathfrak{P}\big(t,m,n \big)$ of $ \sOSp (1|4)$ decomposes into $\mathfrak{so}(3,2)$ subrepresentations as follows
\begin{align}
\mathfrak{P}\big(t,m,n \big)=P\big(t-1,m,n\big) \oplus P\big(t-1,m+1,n\big)  \oplus P\big(t,m+1,n\big) \oplus P\big(t,m+1,n+1\big)~.
\end{align}
For integer ($m=n+1=s$) and half-integer ($m=n=s$) superspin, this decomposition is in agreement with the results of \cite{G-SHR}.


\subsection{Massless supermultiplets} 

To study the component structure of an on-shell  massless supermultiplet of superspin $s>0$, 
it is advantageous to work with the gauge-invariant field strength $\mathfrak{W}_{\a(2s)}$ 
defined by  \eqref{CFS4.8} instead of the prepotential $\f_{\a(m) \ad(n)}$, 
where $2s = m+n+1$. The fundamental properties of $\mathfrak{W}_{\a(2s)}$,
\bea \label{5.16}
\bar \cD_\bd \mathfrak{W}_{\a(2s)}=0~, \qquad
\mathcal{D}^{\b}\mathfrak{W}_{\b \a(2s-1)} =0
\quad \Leftrightarrow \quad \big( \cD^2 - 4(s+1) \bar \m \big) \mathfrak{W}_{\a(2s)} =0~,
\eea
imply the following equations:  
\begin{subequations} \label{5.17}
\bea
\mathcal{D}^{\b \bd}\mathfrak{W}_{\b \a(2s-1)} &=&0~,\\
\big(\mathbb{Q} -  
\l_{(1,2s-1,0)} 
\mu\mub\big)\mathfrak{W}_{\a(2s)} &=&0~.
\eea
\end{subequations}

The chiral field strength $\mathfrak{W}_{\a(2s)}$ has two independent component fields
\begin{subequations} 
\bea
C_{\a(2s)} &:=& \mathfrak{W}_{\a(2s)} |~, \\
C_{\a(2s+1)}&:=& \cD_{(\a_1} \mathfrak{W}_{\a_2 \dots \a_{2s+1})} |~.
\eea
\end{subequations} 
Their properties follow from the equations \eqref{5.16} and \eqref{5.17} 
\begin{subequations}
\bea
\nabla^{\b \bd}C_{\b \a(2s-1)} &=&0 \quad \implies \quad 
\big(\mathcal{Q} - 
\t_{(1,2s,0)} 
\m \bar \m\big) C_{\a(2s)} =0~, \\
\nabla^{\b \bd}C_{\b \a(2s)} &=&0 \quad \implies \quad 
\big( \mathcal{Q} - 
\t_{(1,2s+1,0)} 
\m \bar \m  \big) C_{\a(2s+1)} =0~.
\eea 
\end{subequations}
These relations tell us that the component field strengths $C_{\a(2s)}$ and $C_{\a(2s+1)}$ furnish the massless representation 
$\mathfrak{S}\big(s+1,s\big)=D(s+1, s) \oplus  D\big(s+ \frac{3}{2}, s+ \frac{1}{2}\big) $.


\subsection{Wess-Zumino supermultiplet} \label{WZ section}

Various aspects of the Wess-Zumino model in AdS$_4$ were studied in the 1980s,  
see \cite{IS,BreitenF,BFDG,DF,BG1,BG2} and references therein. 
Here we will only discuss its group-theoretic aspects for completeness. 
In superspace, the Wess-Zumino model
is formulated in terms of a chiral scalar superfield $\Phi$ and its  conjugate $\bar{\Phi}$.
Without self-coupling, the model is described by  the action 
\bea
S_{\text{WZ}}[\Phi, \bar{\Phi}] = \int \rd^{4|4} z\, E\, \Big \{ \Phi \bar{\Phi} + \frac{\l}{2} \Phi^2 + \frac{\bar{\l}}{2}\bar{\Phi}^2 \Big \}~, \qquad \cDB_\ad \Phi = 0~.
\label{5.20}
\eea
Here $\l$ is a dimensionless complex parameter.\footnote{This parameter can be made real by applying a redefinition $\F \to \re^{\ri \g } \F$, where $ \g = \bar \g$ is constant.} 
The  equations of motion corresponding to  this model are
\begin{subequations} \label{EoMWZ}
\bea
- \frac{1}{4}\big(\cDB^2-4\m\big)\bar{\Phi} + \l \m \Phi &=&0~, \\
- \frac{1}{4}\big(\cD^2-4\mub\big)\Phi + \bar{\l} \mub \bar{\Phi} &=&0~.
\eea
\end{subequations}
Using \eqref{EoMWZ}, it can be shown that the superfield $\Phi$ satisfies the mass-shell equation
\be \label{MassShellWZ}
\big(\mathbb{Q} +\m \mub -|\l \m|^2\big)\Phi = 0~.
\ee

The on-shell chiral scalar  $\F$ contains 
two independent component fields, which are
\begin{subequations} \label{CompFields}
\bea
\varphi :&=& \Phi |~, \\
\psi_\a :&=& \cD_\a \Phi|~.
\eea
\end{subequations}
By making use of the superfield mass-shell equation \eqref{MassShellWZ}, it can be shown that the component fields \eqref{CompFields} satisfy the mass-shell equations
\begin{subequations} \label{MassShellWZComp}
\bea
\big(\mathcal{Q}-\r_{(1)}^2 \big)\big(\mathcal{Q}-\r_{(2)}^2 \big)\varphi &=& 0~,\label{MassShellWZComp-a}\\
\big(\mathcal{Q}-\r_{(3)}^2\big)\psi_\a &=& 0~,
\eea
\end{subequations}
where the pseudo-masses $\r_{(i)}^2 $ are 
\begin{subequations} \label{psuedomassWZ}
\bea
\r_{(1)}^2 &=&|\l \m |^2 - \m \mub \big(|\l| + 2\big) ~, \\
\r_{(2)}^2 &=& |\l \m |^2 + \m \mub \big(|\l| - 2\big)~, \\
\r_{(3)}^2 &=& | \l \m |^2 - \frac{3}{2} \m \mub~.
\eea
\end{subequations}
In the massless limit $\l \rightarrow 0$, the mass-shell conditions \eqref{psuedomassWZ} take the form
\begin{subequations} 
	\bea
	\r_{(1)}^2 &=& \r_{(2)}^2 = - 2\mu \mub = \t_{(1,0,0)}\m \mub  ~, \\
	\r_{(3)}^2 &=&- \frac{3}{2} \m \mub = \t_{(1,1,0)}\m \mub~,
	\eea
\end{subequations}
which is consistent with the on-shell massless field conditions \eqref{2.17}. We wish to show that the model \eqref{5.20} describes two
Wess-Zumino representations 
\eqref{E2}. First, let us point out that the equation for $\varphi$, 
\eqref{MassShellWZComp-a},
 is factorised 
into the product of two second-order differential equations
and, hence, describes two spin-$0$ modes with masses $\r_{(1)}$ and $\r_{(2)}$. This means that the Wess-Zumino model
describes three representations of $\mathfrak{so}(3,2)$.
To see that the structure of these representations is consistent with~\eqref{E2}
it is necessary compute the 
minimal energies $E^{(i)}_0$ associated with the pseudo-masses $\r_{(i)}$. Using the prescription advocated in section \ref{section2.2}, we find 
\begin{subequations}
\bea
\big (E_0^{(1)} \big )_\pm &=&\frac{3}{2}\pm \hf \big(2|\l|-1\big) ~, \\
\big (E_0^{(2)} \big )_\pm &=& \frac{3}{2}\pm \hf \big(1+2|\l|\big)~, \\
\big (E_0^{(3)} \big )_\pm &=& \frac{3}{2}\pm |\l|~.
\eea
\end{subequations}
We see that there exist two branches of minimal energy solutions which furnish 
two Wess-Zumino representations $\mathfrak{S}\big(E_0,0\big)$, eq.  \eqref{E2}. 
We will call these branches the positive branch and the negative branch.  
These solutions are given in table \ref{table 4}. 
\begin{table}[h]
	\begin{center}
		\begin{tabular}{|c|c|c|}
			\hline
			~ Component fields  & $E_0$ (positive branch)   & $E_0$ (negative branch)   \Tstrut\Bstrut\\ \hline
\phantom{\Big|} 		spin-$0$       &  $1+|\l|$    & $1 - |\l|$    \Tstrut\\ 
\phantom{\Big|} 	spin-$0$  &  $2 +|\l|$  &  $2-|\l|$   \\ 
\phantom{\Big|} 		spin-$\hf$     &  $\frac{3}{2} + |\l|$   & $\frac{3}{2} - |\l|$  
			\\ 
			\hline
		\end{tabular}
	\end{center} 
	\vspace{-15pt} 
	\caption{Two branches of solutions.}
	\label{table 4}
\end{table}
These branches furnish the Wess-Zumino representations $\mathfrak{S}\big(1+|\l|,0\big)$ and $\mathfrak{S}\big(1-|\l|,0\big)$, eq. \eqref{E2}. 
In order for these representations to be unitary, we require $E_0 > \hf$. 
Therefore, the positive branch describes unitary representations for all values of $\l$. 
The negative branch describes unitary representations for  $|\l| < \hf$. 
In the massless case $\l=0$ the two branches coincide and we have two massless scalars with energies $E_0=1$ and $E_0=2$ and a massless fermion 
with energy $E_0= \frac{3}{2}$.\footnote{As pointed out in section \ref{section rep}, the massless spin-$0$ representations of $\mathfrak{osp}(1|4)$ correspond to two 
possible values of the minimal energy $E_0=1$ or $E_0=2$, 
and the massless spin-$\frac{1}{2}$ representation corresponds to
$E_0= \frac{3}{2}$.}  

There are three special values of $\l$. The choice $\l=0$ corresponds to the superconformal or massless model. For $\l=1$ the model has a dual formulation in terms of a tensor supermultiplet \cite{Siegel-tensor}. In this case the equations of motion 
\eqref{EoMWZ} can be recast in term of a real superfield $L=\F + \bar \F$ 
to take form 
\begin{subequations}\label{tensor}
\bea
\big(\bar \cD^2 - 4\m \big) L = \big( \cD^2 - 4 \bar \m \big) L = 
0~, \label{tensor-a}
\eea
while the chirality of $\F$ gives 
\bea
\big(\bar \cD^2 - 4\m \big) \cD_\a L =0~. \label{tensor-b}
\eea
\end{subequations}
The equations \eqref{tensor} imply that $L$ satisfies the mass-shell equation
\bea
\mathbb{Q}L=0~, \label{tensor2}
\eea
which together define an on-shell scalar superfield. 
Finally for $\l =1/2$ the on-shell chiral scalar $\F$ and its conjugate $\bar \F$ contain 
a sub-representation describing 
the super-singleton, as follows from table \ref{table 4}.


\subsection{Massive vector multiplet} \label{section5.5}

The off-shell massive vector multiplet in a supergravity background was formulated in
\cite{Siegel-tensor,VanProeyen}. In the superspace setting, it is naturally 
 described in terms of a real scalar prepotential $V$.  
 Its action functional  in AdS$^{4|4}$ is given by 
\bea
S [V]= \hf \int\rd^{4|4}z\, E \, V \Big\{ \frac{1}{8}
 \cD^\a \big(\bar \cD^2-4\m\big) \cD_\a + M^2 \Big\} V~, \qquad \bar V = V~,
 \label{vmaction}
 \eea
 with $M$ a non-zero real parameter. 
The corresponding equation of motion 
\bea
\Big( \frac{1}{8}
 \cD^\a \big(\bar \cD^2-4\m\big) \cD_\a + M^2 \Big) V =0
 \label{vmeom-0}
 \eea
implies that $V$ is a linear superfield, 
\begin{subequations}\label{vmeom}
\bea
\big(\bar \cD^2 -4\m\big)  V = \big( \cD^2 -4\bar \m\big)  V=0~.
\label{5.30} 
\eea
Then making use of \eqref{A.4a} gives 
\bea 
\big( \Box +2 \m \bar \m - M^2 \big) V = \big({\mathbb Q} - M^2 \big)V =0 ~.
\label{5.31}
\eea
\end{subequations}
The equations \eqref{5.30} and \eqref{5.31} define an on-shell irreducible supermultiplet.\footnote{It is instructive to compare the system of equations \eqref{tensor} with \eqref{vmeom-0} and its corollaries \eqref{vmeom}. The former describes a non-conformal tensor multiplet in AdS$_4$, which reduces to the free massless tensor multiplet in the flat-superspace limit.}
It is worth pointing out that the above model has a dual formulation, which is the massive tensor multiplet model \cite{Siegel-tensor}.

On the mass shell, the independent component fields of $V$ are:
\begin{subequations}
\bea
A &=& V|~, \\
\j_\a &=& \cD_\a V|~,
\\
h_{\a\ad} &=& \frac{1}{2}\big[ \cD_\a , \bar \cD_\ad \big] V| ~, \qquad 
\nabla^{\a\ad} h_{\a\ad} =0~.
\eea
\end{subequations}
The corresponding equations of motion are:
 \begin{subequations}
\bea
0&=&\big(\cQ+2\m \mub -M^2\big)A~, \label{5.34a} \\
0&=&\big(\cQ+\frac{3}{2}\m \mub -M^2\big)\j_\a - \ri \mub \nabla_{\a}{}^{\bd}\bar{\j}_\bd~, \label{5.34b} \\
0&=&\big(\cQ+\frac{3}{2}\m \mub -M^2\big)\bar{\j}_\ad + \ri \mu \nabla^{\b}{}_{\ad}\j_\b~,
\label{5.34c}\\
0&=&\big(\cQ-M^2\big)h_{\a\ad}~.
\label{5.34d}
\eea
\end{subequations}        
The equations \eqref{5.34b} and \eqref{5.34c} lead to the quartic equation
\bea
0&=&\big(\cQ+\frac{3}{2}\m \mub - M_{+}^2\big)
\big(\cQ+\frac{3}{2}\m \mub - M_{-}^2 \big)\j_\a~, \label{VMquartic}
\eea
where
\bea
M_{\pm}^2:=M^2+ \hf \m \mub \pm \hf \sqrt{\m\mub(4M^2+\m\mub)}
=\frac{1}{4} \Big (\sqrt{\m \mub} \pm \sqrt{4M^2+\m\mub}\Big )^2~.
\eea
The above results imply that the component fields furnish the massive superspin-$\hf$ representation of $\mathfrak{osp}(1,4)$
\bea
\mathfrak{S}\big({E}_0 ,\hf \big)=
D\big(E_0+ \frac{1}{2}, 0\big)  \oplus D\big( E_0, \hf \big) \oplus  
D \big( E_0+1, \hf \big) \oplus  D \big(E_0+ \hf, 1\big) ~, 
\eea
where
$E_0  =1  + \frac{1}{2}
\sqrt{ 4\frac{M^2}{\mu\mub} +1} $.

The spinor equation \eqref{5.34b} and its conjugate \eqref{5.34c} are second-order partial differential equations.
First-order equations for the component spinor fields can be obtained 
if one makes use of a   
St\"uckelberg reformulation of the above model. It is obtained through the replacement 
\begin{align}
V\, \rightarrow \, V+\frac{1}{M}\big(\Phi+\bar{\Phi}\big)~,\qquad \bar{\mathcal{D}}_{\ad}\Phi=0~,
\end{align}
for some chiral superfield $\Phi$. This leads to the action
\bea
S [V,\Phi,\bar{\Phi}]= \hf \int\rd^{4|4}z\, E \, \Big\{  \frac{1}{8}V
 \cD^\a \big(\bar \cD^2-4\m\big) \cD_\a V + \big(MV+\Phi+\bar{\Phi}\big)^2 \Big\} ~,
\eea
which is invariant under gauge transformations of the form
\begin{align}
\delta_{\Lambda}V=\Lambda+\bar{\Lambda}~,\qquad \delta_{\Lambda}\Phi=-M\Lambda~,\qquad \bar{\mathcal{D}}_{\ad}\Lambda=0~. \label{VMGF}
\end{align} 
The corresponding equations of motion are given by 
\begin{subequations}\label{VM}
\begin{align}
0&=\phantom{-}\frac{1}{8}\mathcal{D}^{\a}\big(\bar{\mathcal{D}}^2-4\mu\big)\mathcal{D}_{\a}V +M^2 V + M\big(\Phi+\bar{\Phi}\big)~, \label{VMa} \\
0&= -\frac{1}{4}\big(\bar{\mathcal{D}}^2-4\mu\big)V+\frac{\mu}{M}\Phi-\frac{1}{4M}\big(\bar{\mathcal{D}}^2-4\mu\big)\bar{\Phi}~, \label{VMb}\\
0&= -\frac{1}{4}\big(\mathcal{D}^2-4\mub\big)V+\frac{\mub}{M}\bar{\Phi}-\frac{1}{4M}\big(\mathcal{D}^2-4\mub\big)\Phi~. \label{VMc}
\end{align}
\end{subequations}

Using the gauge freedom \eqref{VMGF}, one may impose the gauge condition $\Phi=0$ whereupon the original model \eqref{vmaction} is recovered. On the other hand, one can instead choose the following Wess-Zumino gauge
\begin{align}
V|=0~, \qquad \mathcal{D}_{\a}V|&=0~,\qquad \mathcal{D}^2V|=0~,\qquad \big(\Phi-\bar{\Phi}\big)|=0~, 
\end{align}
which exhausts the gauge freedom. The remaining non-zero component fields are
\begin{subequations}
\begin{align}
A&=\Phi|~,\\
\psi_{\a}&=\mathcal{D}_{\a}\Phi |~, \label{5.43a}\\
h_{\a\ad}&=\frac{1}{2}\big[\mathcal{D}_{\a},\bar{\mathcal{D}}_{\ad}\big]V|~, \qquad \nabla^{\a\ad}h_{\a\ad}=0~, \label{5.43c} \\
\chi_{\a}&=-\frac{1}{4}\big(\bar{\mathcal{D}}^2-4\mu\big)\mathcal{D}_{\a}V|~.
\end{align}
\end{subequations}
The spinor fields $\psi_{\a}$ and $\chi_{\a}$ and their conjugates 
prove to satisfy the following 
 first-order differential equations
\begin{subequations} \label{FOspin}
\bea
\text{i}\nabla^{\a}{}_{\ad}\chi_{\a}+M\bar{\psi}_{\ad}=0~,  \quad && \quad 
-\text{i}\nabla_{\a}{}^{\ad}\bar \chi_{\ad}+M{\psi}_{\a}=0~,
\label{FOspin-a}\\ 
\text{i}\nabla^{\a}{}_{\ad}\psi_{\a}+M\bar{\chi}_{\ad}+\mub\bar{\psi}_{\ad}=0~,
 \quad && \quad 
 -\text{i}\nabla_{\a}{}^{\ad}\bar{\psi}_{\ad}+M{\chi}_{\a}+\mu {\psi}_{\a}=0~.
 \label{FOspin-b}
\eea
\end{subequations}
Making use of the equations \eqref{FOspin-b} allows us to express 
the fields $\c_\a$ and $\bar \c_\ad$ in terms of $\j_\a$ and $\bar \j_\ad$.
Then the latter fields prove 
to satisfy the equations \eqref{5.34b} and \eqref{5.34c}. 
As regards the bosonic fields \eqref{5.43a} and \eqref{5.43c}, they may be seen to obey the equations \eqref{5.34a} and \eqref{5.34d}. 


\section{Massive gravitino supermultiplet}

As an illustration of our general discussion in section \ref{section 3}, here we present an off-shell  model for the massive gravitino (superspin-1) multiplet in AdS$_4$, 
\bea
\mathfrak{S}\big({E}_0,1\big):=
D\Big(E_0+ \frac{1}{2}, \hf \Big)  \oplus &D( E_0, 1) \oplus  D (E_0+1, 1) \oplus  D\Big(E_0+ \frac{1}{2}, \frac{3}{2}\Big)~.
\label{gravitinorep}
\eea
The unitarity bound for the superspin-1 case is $E_0 \geq 2$, with the  $E_0 =2$ value corresponding to the massless gravitino multiplet. We point out that on-shell models 
(i.e. without auxiliary fields) for the massive gravitino multiplet in AdS$_4$ have appeared in the literature \cite{AB}
(see also \cite{Zinoviev07}).

Two off-shell formulations for the massless gravitino in AdS$_4$ were introduced in \cite{KS94}. One of them  is described by the action\footnote{This model has several dual versions given in \cite{KS94,Butter:2011ym, BHK}. In the flat-superspace limit, the action \eqref{gravitinoaction} reduces to that derived in \cite{GS}.}
\bea \label{gravitinoaction}
&&S_{\text{massless}}[H,\J, \bar \J] = - \int \rd^{4|4}z\, E\,\Big 
\{ \frac{1}{16}H\cD^\a \big (\cDB^2-4\mu \big )\cD_\a H+\m \mub H^2
 \non \\
&& \qquad
+\frac{1}{4}H\big (\cD^\a\cDB^\ad G_{\a\ad} - \cDB^\ad \cD^\a \bar{G}_{\a\ad} \big) 
+\bar{G}^{\a\ad}G_{\a\ad}+\frac{1}{4}\big (\bar{G}_{\a\ad}\bar{G}^{\a\ad}+ G^{\a\ad}G_{\a\ad} \big ) \Big \} ~,  
\eea
where $H$ is a real scalar superfield, 
and $G_{\a\ad}$ is a longitudinal linear superfield constructed in terms of 
an unconstrained prepotential $\J_\a$, 
\bea
G_{\a\ad} = \cDB_\ad \J_\a~, \qquad \bar{G}_{\a\ad} = -\cD_\a \bar{\J}_\ad~.
\eea
We propose a massive extension of the action \eqref{gravitinoaction} by adding a mass term $S_{m}[H,\J, \bar \J] $
\bea
 \label{MassiveAct}
S_{\text{massive}}[H,\J, \bar \J ]=S_{\text{massless}}[H,\J, \bar \J ] 
+ S_{m}[H,\J, \bar \J]~.
\eea
We propose the following mass term:
\bea
S_{m}[H,\J, \bar \J] &=& m \int \rd^{4|4}z\,E\,\Big \{ \J^\a \J_\a + \bar{\J}_\ad\bar{\J}^\ad  +\hf H\big (\cD^\a\J_\a + \cDB_\ad\bar{\J}^\ad \big ) \\
&&- \frac{1}{4}\big(m+2\m + 2\mub\big) H^2 \Big \}~, \non
\eea
where $m$ is a real parameter of unit mass dimension. 

The equations of motion corresponding to $S_{\text{massive}}[H,\J, \bar \J]$,  eq. \eqref{MassiveAct},  are given by
\begin{subequations} \label{EoM}
\bea
0&=&\hf \cD^\a W_\a + \frac{1}{4}\big (\cD^\a \cDB^2\J_\a + \cDB_\ad \cD^2  \bar{\J}^\ad \big) + \frac{m}{2} \big ( \cD^\a \J_\a + \cDB_\ad \bar{\J}^\ad \big ) \label{EoM1}\\
&&-\big( m \m + m \mub +2 \m\mub+\hf m^2\big)H ~, \non \\
0 &=&-W_\a+ \big(\frac{m}{2}+\m\big)\cD_\a H + \cDB_\ad \cD_\a \bar{\J}^\ad -\hf \cDB^2 \J_\a -2m \J_\a ~, \label{EoM2}
\eea
\end{subequations}
where we have introduced the field strength $W_\a = -\frac{1}{4}\big(\cDB^2-4\m\big)\cD_\a H$. Acting on \eqref{EoM1} and \eqref{EoM2} with $\cDB^2$ gives
\begin{subequations} \label{DBarSquareCond}
\bea 
\big(\cDB^2 - 4\m\big) \cD^\a \J_\a &=&  \big(m+2\mub\big) \big(\cDB^2 - 4\m\big) H~, \\
\big(\cDB^2 -4\m\big)\J_\a &=& - W_\a~.
\eea
\end{subequations}
Next, acting on \eqref{EoM2} with $\cD^\a$ yields
\bea \label{DCond}
\cDB_\ad\bar{\J}^\ad = \big(m+2\m\big)H-\frac{1}{\mub}\big(m+\m\big)\cD^\a\J_\a~.
\eea
Using relations \eqref{EoM1}, \eqref{DBarSquareCond} and \eqref{DCond}, 
one can show that 
\be \label{TransverseCondGrav}
\cD^\a \J_\a =0~.
\ee
Substituting \eqref{TransverseCondGrav}  into \eqref{DCond}, it immediately follows that 
\bea
H=0
\eea
on the mass shell. 
Now, the above relations imply that the spinor $\J_\a$ obeys 
the following mass-shell conditions:
\begin{subequations}
\bea
\cD^\a \J_\a &=& 0~, \quad \big(\cDB^2 -4\mu\big)\J_\a = 0~, \\
\ri \cD_{\a\ad} \bar{\J}^\ad + \big(m+\m\big) \J_\a &=&0 \quad \Longrightarrow \quad \big (\mathbb{Q}-(m+\mu)(m+\mub) - \m \mub  \big )\J_\a =0~. ~~~
\label{MassShellCond}
\eea
\end{subequations}
These are the on-shell irreducibility constraints for a massive superspin-1 multiplet, as defined in section \ref{section 3}. The unitarity bound for the massive gravitino multiplet is 
\bea
|m +\m |^2 > \m \bar \m  \quad \implies \quad m +2\m \neq 0~.
\eea
These conditions have been used in the above derivation. One may think of the 
first Dirac-type equation in \eqref{MassShellCond} as the  reality condition relating 
$\J_\a$ and $\bar \J_\ad$.

In conclusion, we give the explicit expression for 
the massive gravitino action \eqref{MassiveAct}
\bea
&&S_{\text{massive}}[H,\J, \bar \J ] =\int \rd^{4|4}z\, E\,\Big \{ -\frac{1}{16}H\cD^\a \big (\cDB^2-4\mu \big )\cD_\a H-\m \mub H^2
~ \non \\
&&\qquad  +\frac{1}{4}H\big (\cD^\a\cDB^2 \J_\a + \cDB_\ad \cD^2 \bar{\J}^\ad \big )
+\cD^\a \bar{\J}^\ad \cDB_{\ad}\J_\a
+\frac{1}{4}\big (\bar{\J}_{\ad}\cD^2\bar{\J}^{\ad}+ \J^{\a}\cDB^2 \J_\a \big )  ~ \non
\\
&&\qquad - \frac{m}{4}\big(m+2\m +2\mub\big) H^2 +\frac{m}{2} H\big (\cD^\a\J_\a + \cDB_\ad\bar{\J}^\ad \big )+ m \big(\J^\a \J_\a +  \bar{\J}_\ad \bar{\J}^\ad\big)
 \Big \}~.~~~~~~
\eea
Note that in the flat superspace limit, $\m \rightarrow 0$, the action reduces to that describing the massive gravitino model derived in \cite{BGKP}. 

In the massless limit, $m \rightarrow 0$, the second mass-shell condition \eqref{MassShellCond} becomes 
\be
\big(\mathbb{Q}-2\mu \mub \big)\J_\a = \big(\mathbb{Q}-\l_{(1,1,0)}\m \mub\big)\J_\a = 0~.
\ee
This agrees with the definition of on-shell massless supermultiplets 
given in section  \ref{section 3}.

On the mass shell, the independent component fields of $\J_\a$ are:
\begin{subequations}\label{GravComp}
\bea
\c_\a &=& \J_\a|~,\\
A_{\a\ad}&=&\cDB_\ad \J_\a| ~, \qquad \nabla^{\a\ad} A_{\a\ad}=0~,\\
B_{\a\b}&=& \cD_{(\a}\J_{\b)}|~,\\
\varphi_{\a\b\ad}&=&\Big (\hf [\cD_{(\a},\cDB_{\ad} ] - \frac{\ri}{3}\mathcal{D}_{(\a\ad} \Big )\J_{\b)}|~, \qquad \nabla^{\a\ad}\varphi_{\a\b\ad}=0~.
\eea
\end{subequations}
Here $A_{\a\ad}$ is a complex vector field.
The component equations of motion, which
 follow from the first equation in \eqref{MassShellCond}, are:
\begin{subequations}\label{6.16}
\bea
0&=&\ri \nabla^\b{}_{\ad}\c_\b +\big(m+\mub\big)\bar{\c}_\ad~, \\
0&=& \ri \nabla^\b{}_{\ad} B_{\a\b}+\mub A_{\a\ad}-\big(m+\mub\big)\bar{A}_{\a\ad}~, \\
0&=&\ri \nabla^\b{}_{(\ad}A_{\b\bd)}-\big(m+\mub\big)\bar{B}_{\ad\bd}~, \\
0&=&\ri \nabla^\b{}_{(\ad}\varphi_{\a\b\bd)}+\big(m+\mub\big)\bar{\varphi}_{\a\ad\bd}~.
\eea
\end{subequations}
These equations can be viewed as reality conditions expressing the component fields of 
$\bar \J_\ad$ in terms of those contained in $\J_\a$. 
Making use of the relations  \eqref{6.16} and their conjugates
leads to the following equations: 
\begin{subequations}
\bea
0&=&\big (\cQ+\frac{3}{2}\m\mub-\k^2 \big )\c_\a~,\\
0&=&\big (\cQ-\k^2\big )B_{\a\b}-\ri \mub \nabla_{(\a}{}^\bd A_{\b)\bd}~, \label{EntG1}\\
0&=&\big (\cQ+\m \mub -\k^2 \big ) A_{\a\ad}+\ri \m \nabla^\b{}_\ad B_{\a\b}~, \label{EntG2} \\
0&=&\big (\cQ-\frac{3}{2}\m \mub -\k^2 \big )\varphi_{\a\b\ad}~,
\eea
\end{subequations}
where
\be
\k^2:=|m+\m |^2~.
\ee
It can be shown that the equations \eqref{EntG1} and \eqref{EntG2} imply the quartic equation
\begin{subequations}
\bea
0&=&\big (\cQ-\k^2-\sqrt{\m \mub}\k \big ) \big (\cQ-\k^2+\sqrt{\m \mub}\k \big )A_{\a\ad}~,\\
0&=&\big (\cQ-\k^2-\sqrt{\m \mub}\k \big )\big (\cQ-\k^2+\sqrt{\m \mub}\k \big )B_{\a\b}~.
\eea
\end{subequations}
The analysis above indicates that the component fields furnish the massive superspin-$1$ representation  \eqref{gravitinorep}, where $E_0=1+\sqrt{\frac{\k^2}{\m \mub}}$. Additionally, the mass-shell conditions are consistent with the results in section 5.1, for a superfield with index structure $m-1=n=0$, upon the redefinition $\k^2=M^2-\mu \mub$.


\section{Factorisation of superconformal higher-spin
actions} \label{section 4}

In section \ref{construction} we introduced the gauge invariant actions for superconformal higher-spin multiplets and used them to motivate our ansatz for the TLAL projectors. Let us come back to this topic and demonstrate that these actions factorise into products of minimal second order differential operators. 

We begin with the familiar case when $m=n=s$ and impose the reality condition \eqref{reality}. It is clear that for these values the higher-spin Bach tensor \eqref{Bach1} and the descendant \eqref{P1} coincide, so that we have the relations
\begin{align}
\mathfrak{B}_{\a(s)\ad(s)}(H)=\mathbb{P}_{\a(s)\ad(s)}(H)= \prod_{t=1}^{s+1} \big ( \mathbb{Q}-\l_{(t,s,s)}\m \mub \big )\P_{\a(s)\ad(s)}(H)~.\label{bosbach}
\end{align}
Now, one can employ the decomposition from section \ref{decomp} (which is valid for all off-shell superfields) to split the unconstrained prepotential $H_{\a(s)\ad(s)}$ in the action \eqref{SCHS1} into TLAL and longitudinal components,
\begin{align}
H_{\a(s)\ad(s)}={\mathfrak H}_{\a(s)\ad(s)}+\bar{\mathcal{D}}_{(\ad_1}\z_{\a(s)\ad_2\dots\ad_s)}-\mathcal{D}_{(\a_1}\bar{\z}_{\a_2\dots\a_s)\ad(s)}~.
\end{align}
Here  ${\mathfrak H}_{\a(s)\ad(s)}$ is TLAL, whilst $\z_{\a(s)\ad(s-1)}$ is complex and unconstrained. 
 
 By virtue of the gauge invariant and TLAL nature of the higher-spin Bach tensor, the longitudinal parts give vanishing contribution to the action since they correspond to the pure gauge sector. Therefore only the TLAL component of the prepotential remains and, since the projector $\Pi^{(s,s)}$ acts as the identity on this subspace, using \eqref{bosbach} we obtain the following factorisation
\begin{align}
 S_{\text{SCHS}}^{(s,s)}[H]=2(-1)^s\int\text{d}^{4|4}z \, E \, {\mathfrak H}^{\a(s)\ad(s)} \prod_{t=1}^{s+1} \big ( \mathbb{Q}-\l_{(t,s,s)}\m \mub \big ){\mathfrak H}_{\a(s)\ad(s)} ~. \label{fact1}
\end{align}
 This process is equivalent to fixing the gauge freedom \eqref{GT} by imposing the gauge condition $H_{\a(s)\ad(s)}\equiv{\mathfrak H}_{\a(s)\ad(s)} $, since the action \eqref{fact1} no longer possesses any gauge symmetry. 

Next we consider the complex supermultiplet $\Phi_{\a(m)\ad(n)}$ with $n > m$. In contrast to \eqref{bosbach}, one may show that in this case the following relation holds
\begin{align}
\mathbb{P}_{\a(m)\ad(n)}(\Phi)=\mathcal{D}_{(\ad_1}{}^{\b_1}\cdots\mathcal{D}_{\ad_{n-m}}{}^{\b_{n-m}}\mathfrak{B}_{\a(m)\b(n-m)\ad_{n-m+1}\dots\ad_{n})}(\Phi)~,
\end{align}
which upon inverting yields
\begin{align}
\mathfrak{B}_{\a(n)\ad(m)}(\Phi)= \bigg [ \prod_{t=m+2}^{n+1} \big ( \mathbb{Q}-\l_{(t,m,n)}\m \mub \big ) \bigg ]^{-1}&\mathcal{D}_{(\a_1}{}^{\bd_1}\cdots\mathcal{D}_{\a_{n-m}}{}^{\bd_{n-m}} \non\\
&\times \mathbb{P}_{\a_{n-m+1}\dots\a_n)\bd(n-m)\ad(m)}(\Phi)~.
\end{align}
In terms of the TLAL projector this reads
\begin{align}
\mathfrak{B}_{\a(n)\ad(m)}(\Phi)=\prod_{t=1}^{m+1} \big ( \mathbb{Q}-\l_{(t,m,n)}\m \mub \big ) &\mathcal{D}_{(\a_1}{}^{\bd_1}\cdots\mathcal{D}_{\a_{n-m}}{}^{\bd_{n-m}}\non\\
&\times \Pi_{\a_{n-m+1}\dots\a_n)\bd(n-m)\ad(m)}(\Phi) ~.
\end{align}
Employing the same trick as in the previous case, one obtains the following factorisation of the SCHS action \eqref{SCHS1} 
\begin{align}
 S_{\text{SCHS}}^{(m,n)}[\Phi,\bar{\Phi}]=\text{i}^{m+n}&\int\text{d}^{4|4}z \, E \, \bar{\phi}^{\a(n)\ad(m)}\prod_{t=1}^{m+1} \big ( \mathbb{Q}-\l_{(t,m,n)}\m \mub \big ) \non\\
 &\times\mathcal{D}_{(\a_1}{}^{\bd_1}\cdots\mathcal{D}_{\a_{n-m}}{}^{\bd_{n-m}}\phi_{\a_{n-m+1}\dots\a_n)\bd(n-m)\ad(m)}+\text{c.c.}~, \label{fact2}
\end{align}
where $\phi_{\a(m)\ad(n)}$ is the TLAL part of $\Phi_{\a(m)\ad(n)}$. We see that, due to the mismatch of $m$ and $n$, the SCHS action does not factorise wholly into products of second-order operators. However, upon taking appropriate derivatives of the equation of motion resulting from varying $\bar{\phi}_{\a(m)\ad(n)}$ in \eqref{fact2}, one arrives at the following fully factorised on-shell equation 
\begin{align}
0=\prod_{t=1}^{n+1} \big ( \mathbb{Q}-\l_{(t,m,n)}\m \mub \big )\phi_{\a(m)\ad(n)}~. \label{eom1}
\end{align}
According to our definitions in section \ref{section 3}, we see that some extra (non-unitary) massive modes, corresponding to the values of $\lambda_{(t,m,n)}$ with $m+1< t \leq n+1$ enter the spectrum of the wave equation \eqref{eom1}.

Finally, if $m > n$, then one may show that the following relation holds
\begin{align}
\widehat{\mathbb{P}}_{\a(m)\ad(n)}(\Phi)=\mathcal{D}_{(\ad_1}{}^{\b_1}\cdots\mathcal{D}_{\ad_{m-n}}{}^{\b_{m-n}}\widehat{\mathfrak{B}}_{\a(n)\b(m-n)\ad_{m-n+1}\dots\ad_{m})}(\Phi)~.
\end{align}
Upon inverting and expressing in terms of the TLAL projector, one may write
\begin{align}
\mathfrak{B}_{\a(n)\ad(m)}(\Phi) 
=\prod_{t=1}^{n+1} \big ( \mathbb{Q}-\l_{(t,m,n)}\m \mub \big )& \mathcal{D}_{(\ad_1}{}^{\b_1}\cdots\mathcal{D}_{\ad_{m-n}}{}^{\b_{m-n}}\non\\
&\times\Pi_{\a(n)\b(m-n)\ad_{m-n+1}\dots\ad_m)}(\Phi) ~,
\end{align}
where we have used the identities \eqref{coincidence} and $\widehat{\mathfrak{B}}_{\a(n)\ad(m)}(\Phi)=\mathfrak{B}_{\a(n)\ad(m)}(\Phi)$. In this case the factorised action takes the form 
\begin{align}
 S_{\text{SCHS}}^{(m,n)}[\Phi,\bar{\Phi}]=\text{i}^{m+n}&\int\text{d}^{4|4}z \, E \, \bar{\phi}^{\a(n)\ad(m)}\prod_{t=1}^{n+1} \big ( \mathbb{Q}-\l_{(t,m,n)}\m \mub \big ) \non\\
 &\times\mathcal{D}_{(\ad_1}{}^{\b_1}\cdots\mathcal{D}_{\ad_{m-n}}{}^{\b_{m-n}}\phi_{\a(n)\b(m-n)\ad_{m-n+1}\dots\ad_m)}+\text{c.c.}~, \label{fact3}
\end{align}
whilst the analogous on-shell equation is
\begin{align}
0=\prod_{t=1}^{m+1} \big ( \mathbb{Q}-\l_{(t,m,n)}\m \mub \big )\phi_{\a(m)\ad(n)}~. \label{eom2}
\end{align}
Once again, (non-unitary) massive modes appear in the resulting wave equation. 

In the non-supersymmetric case, the factorisation of the conformal operators was observed long ago in \cite{DeserN1,Tseytlin5, Tseytlin6} for the lower-spin values $s=3/2 $ and $s= 2$. The factorisation of the higher-spin conformal operators was conjectured in \cite{Tseytlin13,Karapet1} (see also \cite{Karapet2}), and later proved by several groups \cite{Metsaev2014, NT, GH, KP20}. In this section we have provided the first derivation of the factorisation of the superconformal higher-spin actions in $\mathcal{N}=1$ superspace. Just as it is for the non-supersymmetric case, the latter factor into products of minimal second-order differential operators involving all partial mass values. 

\section{Conclusion} \label{Discussion}



In conclusion we briefly summarise the main results of this paper. 
We constructed the superspin projection operators 
$\Pi^{(m,n)}$ in AdS${}^{4|4}$.
They are given by eq. \eqref{projectors} for $m\geq n>0$, 
and by eq. \eqref{projectors-m} for $m>n=0$ (the other cases are obtained by complex conjugation). 
The operator  $\Pi^{(m,n)}$ maps an unconstrained superfield $\Phi_{\a(m)\ad(n)}$ into one possessing the properties of a conserved conformal supercurrent, see the equations \eqref{TLAproj} and \eqref{3.34}. 
Making use of the superprojectors, 
we obtained a new representation for the superconformal higher-spin gauge actions in AdS$_4$ given by  eq. \eqref{3.40}, which was used to demonstrate their factorisation, as discussed in section \ref{section 4}.

We provided a systematic discussion of how to realise the unitary (both massive and massless) and the partially massless representations of the ${\cal N}=1$ AdS${}_4$ superalgebra $\mathfrak{osp} (1|4)$ in terms of on-shell superfields. 
In particular, we established a one-to-one correspondence between 
the on-shell partially massless supermultiplets $\phi_{\a(m)\ad(n)}$ with super-depth $t$ and the poles of $\Pi^{(m,n)}$ determined by  $\lambda_{(t,m,n)}\mu\mub$ 
 belonging to  the range \eqref{MS}. 
 The corresponding gauge transformations were derived in the $m=n$ case. 

Our results make it possible to address a number of interesting open problems including the computation of partition functions for the superconformal higher-spin gauge theories in AdS$_4$. This will be discussed elsewhere. 
\\

\noindent
{\bf Acknowledgements:}\\
SMK is grateful to I. L. Buchbinder and A. A. Tseytlin for email correspondence. 
The work of EIB and SMK is supported in part by the Australian 
Research Council, project No. DP200101944.
The work of DH is supported by the Jean Rogerson Postgraduate Scholarship and an Australian Government Research Training Program Scholarship.  
The work of MP is supported by the Hackett Postgraduate Scholarship UWA,
under the Australian Government Research Training Program.


\appendix

\section{AdS superspace toolkit} \label{appendixA}

Our two-component spinor notation and conventions follow \cite{BK}.
In particular, the Lorentz generators $M_{\a\b} $ and ${\bar M}_{\ad \bd}$ 
act on two-component spinors as follows: 
\begin{subequations} 
\bea
M_{\a\b} \,\j_\g=
\hf(\ve_{\g\a}\j_{\b}+\ve_{\g\b}\j_{\a})
~,\quad&&\qquad M_{\a\b}\, {\bar \j}_{\gd}=0~,\\
{\bar M}_{\ad\bd} \,{\bar \j}_{\gd}=
\hf(\ve_{\gd\ad}{\bar \j}_{\bd}+\ve_{\gd\bd}{\bar \j}_{\ad})
~,\quad&&\qquad {\bar M}_{\ad\bd}\, \j_{\g}=0~.
\eea
\end{subequations} 
In this paper we always work with tensor superfields 
 that are symmetric in their undotted indices and separately in the dotted ones,
$\F_{\a(m)\ad(n)} = \F_{(\a_1 ... \a_m)(\ad_1 ... \ad_n)} = \F_{\a_1 ... \a_m\ad_1 ... \ad_n} $.
The following identities hold:
\begin{subequations}
	\bea
	M_{\a_1}{}^{\b}\F_{\b \a_2 ... \a_{m}\ad(n)} &=& - \hf (m+2)\F_{\a(m)\ad(n)}~,\\
	\bar{M}_{\ad_1}{}^{\bd}\F_{\a(m)\bd \ad_2 ... \ad_{n}} &=& - \hf (n+2)\F_{\a(m)\ad(n)}~, \\
	M^{\b\g}M_{\b\g}  \F_{\a(m)\ad(n)} &=& -\hf m(m+2) \F_{\a(m)\ad(n)}~, \\
	\bar{M}^{\bd\gd} \bar{M}_{\bd\gd} \F_{\a(m)\ad(n)} &=& - \hf n(n+2)\F_{\a(m)\ad(n)}~.
	\eea
\end{subequations}

We often make use of the following identities, which can be readily derived from the algebra of covariant derivatives \eqref{algebra}:
\begin{subequations} 
\label{A.2}
\bea 
\cD_\a\cD_\b
\!&=&\!\frac{1}{2}\ve_{\a\b}\cD^2-2{\bar \m}\,M_{\a\b}~,
\quad\qquad \,\,\,
{\bar \cD}_\ad{\bar \cD}_\bd
=-\frac{1}{2}\ve_{\ad\bd}{\bar \cD}^2+2\m\,{\bar M}_{\ad\bd}~,  \label{A.2a}\\
\cD_\a\cD^2
\!&=&\!4 \bar \m \,\cD^\b M_{\a\b} + 4{\bar \m}\,\cD_\a~,
\quad\qquad
\cD^2\cD_\a
=-4\bar \m \,\cD^\b M_{\a\b} - 2\bar \m \, \cD_\a~, \label{A.2b} \\
{\bar \cD}_\ad{\bar \cD}^2
\!&=&\!4 \m \,{\bar \cD}^\bd {\bar M}_{\ad\bd}+ 4\m\, \bar \cD_\ad~,
\quad\qquad
{\bar \cD}^2{\bar \cD}_\ad
=-4 \m \,{\bar \cD}^\bd {\bar M}_{\ad\bd}-2\m\, \bar \cD_\ad~,  \label{A.2c}\\
\left[\bar \cD^2, \cD_\a \right]
\!&=&\!4\rm i \cD_{\a\bd} \bar \cD^\bd +4 \m\,\cD_\a = 
4\rm i \bar \cD^\bd \cD_{\a\bd} -4 \m\,\cD_\a~,
 \label{A.2d} \\
\left[\cD^2,{\bar \cD}_\ad\right]
\!&=&\!-4\rm i \cD_{\b\ad}\cD^\b +4\bar \m\,{\bar \cD}_\ad = 
-4\rm i \cD^\b \cD_{\b\ad} -4 \bar \m\,{\bar \cD}_\ad~,
 \label{A.2e}
\eea
\end{subequations} 
where $\cD^2=\cD^\a\cD_\a$, and ${\bar \cD}^2={\bar \cD}_\ad{\bar \cD}^\ad$. 
Other useful identities are: 
\begin{subequations}\label{A.4}
	\bea
	\cD_\a{}^\bd \cD_\bd{}^\b &=& \d_\a{}^\b \Box -2\m\mub M_\a{}^\b~, \label{A.4a}\\
	\cD^\a{}_\ad \cD_\a{}^\bd &=& \d_\ad{}^\bd \Box - 2 \mu \mub \bar{M}_\ad{}^\bd~.
	\eea
\end{subequations}
Of special importance are the relations:
\begin{subequations} \label{A.5}
\bea
\Box +2\m \bar \m = 
- \frac 18 \mathcal{D}^{\a}\big(\bar{\mathcal{D}}^2-4\mu\big)\mathcal{D}_{\a} 
&+&\frac{1}{16} \big\{\mathcal{D}^2 -4\bar \m ,\bar{\mathcal{D}}^2 -4 \m\big\} ~, \label{A.5a}\\
\mathcal{D}^{\a}\big(\bar{\mathcal{D}}^2-4\mu\big)\mathcal{D}_{\a}
&=& \bar{\mathcal{D}}_{\ad}\big(\mathcal{D}^2-4\mub\big)\bar{\mathcal{D}}^{\ad}~,\\
\big[\mathcal{D}^2,\bar{\mathcal{D}}^2\big]=-4\text{i}\big[\mathcal{D}^\b,\bar{\mathcal{D}}^{\bd}\big]\mathcal{D}_{\b\bd}&+&8\mu\mathcal{D}^2-8\mub\bar{\mathcal{D}}^2 \non \\
=-4\text{i}   \mathcal{D}_{\b\bd}\big[\mathcal{D}^\b,\bar{\mathcal{D}}^{\bd}\big]
&-&8\mu\mathcal{D}^2+8\mub\bar{\mathcal{D}}^2~.
\eea
\end{subequations}

The isometry group of AdS superspace,  AdS$^{4|4}$,  is $\sOSp(1|4)$.
The  isometry transformations of AdS$^{4|4}$ are generated by the Killing supervector fields
$\x^A E_A$ which are defined to solve the Killing equation
\bea
\big[\X,\cD_{A} \big]=0~,\qquad
\X:=-\hf \x^{\b\bd } \cD_{\b\bd}+\x^\b \cD_\b+{\bar \x}_\bd {\bar \cD}^\bd 
+ \x^{\b\g}M_{\b\g} +\bar \x^{\bd \gd} \bar M_{\bd \gd} ~, 
\label{A.6}
\eea
for some Lorentz parameter  $\x^{\b\g}  = \x^{\g\b}$. 
Given a supersymmetric field theory in AdS$_4$ formulated in terms of superfield dynamical variables $\mathfrak V$ (with suppressed indices), its action is invariant under 
the isometry transformations 
\bea
\d {\mathfrak V} = \X  \mathfrak V ~, 
\label{SUSY}
\eea
with $\x^B$ being  an arbitrary Killing supervector field.

The Killing equation \eqref{A.6} implies the following \cite{BK}:
\begin{subequations} \label{A.7}
\bea
\cD_\a \x_{\b\bd} &=& 4\ri \ve_{\a\b} \bar \x_\bd~, \qquad \qquad 
\bar \cD_\ad \x_{\b\bd} = -4\ri \ve_{\ad\bd}  \x_\b~, \\
\cD_\a \x_\b &=&\x_{\a\b} ~,\qquad  \qquad \qquad \bar \cD_\ad \x_\b = -\frac{\ri}{2} \m \x_{\b \ad}~, \\
\cD_\a \x_{\b\g} &=& -4\bar \m \ve_{\a (\b} \x_{\g)}~,  \qquad ~\bar \cD_\ad \x_{\b\g}=0~.
\eea
\end{subequations}
It is seen that the parameters $\x_\b$, $\x_{\b\g}$ and their conjugates are expressed in terms of the vector parameter $\x_{\b \bd}$, and the latter obeys the equation 
\bea
\cD_{(\a}\x_{\b)\bd}=0 \quad \implies  \quad \cD_{(a} \x_{b)}=0~.
\label{A.8} 
\eea
It also follows from \eqref{A.7} that, for every element of the set of parameters $\U =\{ \x^B, \x^{\b\g}, \bar \x^{\bd \gd} \}$, its covariant derivative $\cD_A \U$ is a linear combination of the elements of $\U$. Therefore, all information about the superfield parameters
$\U$ is encoded in their bar-projections, $\U|$. 

Every Killing supervector superfield $\x^B$ on AdS$^{4|4}$, eq. \eqref{A.6}, can be uniquely decomposed as a sum of even and odd ones. The Killing supervector field $\xi^{B}$ is defined to be even if
\bea
\label{even}
v^{b} := \xi^{b}| \neq 0 ~, \quad \xi^\b| = 0~.
\eea
and odd if
\bea
\label{odd}
\xi^{b}|  = 0 ~, \quad \e^{\b} := \xi^{\b}| \neq 0~.
\eea
The fields $v^{b}(x) $ and $\e^{\b}(x) $ encode complete information about the parent conformal Killing vector superfield. It follows from \eqref{A.8} that $v^b$ is a Killing vector field on AdS$_4$, 
\bea
\nabla_{(a} v_{b)}=0~.
\eea
The relations \eqref{A.7} imply that $\e^\b$ is a Killing spinor field satisfying the equation
\bea
\nabla_{\a\ad} \e_\b = \frac{\ri}{4} \m \ve_{\a\b} \bar \e_\ad~.
\eea
Every Killing vector  $v^b$ on AdS$_4$ can be lifted to 
an even  Killing supervector field $\x^B$ on AdS$^{4|4}$ using the relations \eqref{A.7}.
A similar statement holds for Killing spinors.

Given a tensor superfield ${\mathfrak V}$ (with suppressed indices), its independent component fields are contained in the set of fields 
$\vf=\F|$, where  
$\F:=\big\{ {\mathfrak V}, \cD_{ \a} {\mathfrak V}, \bar \cD_{ \ad} {\mathfrak V}, \dots \big\}$.
In accordance with \eqref{SUSY}, the supersymmetry transformation of $\vf$ is 
\bea
\d_\e \vf =  \e^\b (\cD_\b \F)| + \bar \e_\bd (\bar \cD^\bd  \F )|~.
\label{A.14}
\eea


\section{Partially massless gauge symmetry} \label{appendixB}

This appendix is devoted to the derivation of partially massless gauge transformations both in the non-supersymmetric and supersymmetric cases. 


\subsection{The non-supersymmetric case}

Given two integers $m$ and $n$ such that $m\geq n>0$, 
let $h_{\a(m)\ad(n)}$  be an on-shell field,
\begin{subequations}\label{B.1}
\begin{align}
0&=\big(\mathcal{Q}- \rho^2
\big)h_{\a(m)\ad(n)}~,\label{B.1a}\\
0&= \nabla^{\b\bd}h_{\a(m-1)\b\ad(n-1)\bd}~.\label{B.1b}
\end{align}
\end{subequations}
We would like to determine those values of $\rho$ for which the above system of equations
is compatible with a gauge symmetry. 

We begin by positing a gauge transformation of the form 
\begin{align}
\delta_{\zeta}h_{\a(m)\ad(n)}=\nabla_{(\a_1(\ad_1}\zeta_{\a_2\dots\a_m)\ad_2\dots\ad_n)}~ \label{BGT}
\end{align}
and look for a gauge parameter  $\zeta_{\a(m-1)\ad(n-1)}$ such that 
$\delta_{\zeta}h_{\a(m)\ad(n)}$ is a solution to the equations \eqref{B.1}. 
Clearly, for gauge invariance of \eqref{B.1a} the gauge parameter must satisfy
\begin{align}
0&=\big(\mathcal{Q}- \rho^2
\big)\zeta_{\a(m-1)\ad(n-1)}~. \label{Btk}
\end{align}
Next we require \eqref{B.1b} to be gauge invariant, 
$\nabla^{\b\bd} \d_\z h_{\a(m-1)\b\ad(n-1)\bd} =0$.
To solve this problem, let us recall a technical result derived in \cite{KP20}. 
Using the spin projection operators, it was shown in \cite{KP20} that any unconstrained tensor field can be decomposed into irreducible (i.e. transverse) parts.  For the gauge parameter $\zeta_{\a(m-1)\ad(n-1)}$, this decomposition takes the form
\begin{align}
\zeta_{\a(m-1)\ad(n-1)}=~&\zeta_{\a(m-1)\ad(n-1)}^{\text{T}}+\sum_{t=1}^{n-1}\nabla_{(\a_1(\ad_1}\cdots\nabla_{\a_t\ad_t}\zeta^{\text{T}}_{\a_{t+1}\dots\a_{m-1})\ad_{t+1}\dots\ad_{n-1})}~,
\label{Bty}
\end{align} 
where $\zeta^{\text{T}}_{\a(m-n)}$ is unconstrained, whilst the other fields are transverse,
\begin{align}
0=\nabla^{\b\bd}\zeta^{\text{T}}_{\a(m-t-1)\b\ad(n-t-1)\bd}~,\qquad 1 \leq t \leq n-1~. \label{Bth}
\end{align}
Inserting this into the gauge transformations \eqref{BGT} and requiring that the condition \eqref{B.1b} be gauge invariant, one arrives at the following equation\footnote{To derive this equation the identities in the appendix of \cite{KP20} are indispensable.}
\begin{align}
0=\sum_{k=1}^{n}k(m+n-k+1)&\Big(
\rho^2
-\tau_{(k,m,n)} \mu\mub\Big)
\non\\
&\times\nabla_{(\a_1(\ad_1}\cdots\nabla_{\a_{k-1}\ad_{k-1}}\zeta^{\text{T}}_{\a_k\dots\a_{m-1})\ad_k\dots\ad_{n-1})}~.\label{Btx}
\end{align}
The right-hand side of \eqref{Btx} is the decomposition into irreducible parts.
The whole expression may vanish only  if  there exists an integer $t$ such that 
\bea
\rho^2 =  \tau_{(t,m,n)}\mu\mub~, \qquad  1 \leq t \leq n~.
\label{B.7}
\eea
In addition,
each $\zeta^{\text{T}}_{\a(m-k)\ad(n-k)}$ in \eqref{Btx} except for $\zeta^{\text{T}}_{\a(m-t)\ad(n-t)}$ must vanish identically. Hence the decomposition \eqref{Bty} reduces to
\begin{align}
\zeta_{\a(m-1)\ad(n-1)}&=\nabla_{(\a_1(\ad_1}\cdots\nabla_{\a_{t-1}\ad_{t-1}}\zeta^{\text{T}}_{\a_{t}\dots\a_{m-1})\ad_{t}\dots\ad_{n-1})}~, 
\end{align}
and the gauge transformation \eqref{BGT} becomes the  well-known one for a partially-massless field with depth $t$,
\begin{align}
\delta_{\zeta}h_{\a(m)\ad(n)}&=\nabla_{(\a_1(\ad_1}\cdots\nabla_{\a_{t}\ad_{t}}\zeta^{\text{T}}_{\a_{t+1}\dots\a_{m})\ad_{t+1}\dots\ad_{n})}~,
\end{align}
with $\zeta^{\text{T}}_{\a(m-t)\ad(n-t)}$ satisfying \eqref{Btk}, \eqref{Bth} 
and \eqref{B.7}.

\subsection{The supersymmetric case} \label{B.2}

We first consider the real supermultiplet $H_{\a\ad}=\bar{H}_{\a\ad}$ which is on-shell
\begin{subequations} \label{B.35}
\begin{align}
0&=\big(\mathbb{Q}-M^2\big)H_{\a\ad}~,\label{B.35a}\\
0&=\mathcal{D}^{\b}H_{\b\ad}=\bar{\mathcal{D}}^{\bd}H_{\a\bd}~. \label{B.35b}
\end{align}
\end{subequations}
Once again we would like to determine those values of $M$ for which the above system of equations
is compatible with a gauge symmetry. We will see that this occurs only at partially massless values. For the supermultiplet $H_{\a\ad}$ there are only two such values, 
\begin{subequations} \label{B.355}
\begin{align}
M_{(1)}^2&\equiv \lambda_{(1,1,1)}\mu\mub=5\mu\mub~,\label{B.355a}\\
M_{(2)}^2& \equiv \lambda_{(2,1,1)}\mu\mub=3\mu\mub~,\label{B.355b}
\end{align}
\end{subequations}
corresponding to the super-depths $t=1$ (massless) and $t=2$ respectively.

We begin by positing a gauge transformation of the form 
\begin{align}
\delta H_{\a\ad}= \bar{\mathcal{D}}_{\ad}\Lambda_{\a}-\mathcal{D}_{\a}\bar{\Lambda}_{\ad}\label{B.36}
\end{align}
for an unconstrained gauge parameter  $\Lambda_{\a}$. From this we must deduce which constraints must be placed upon $\Lambda_{\a}$ in order for $\delta H_{\a\ad}$ to be a solution to the equations \eqref{B.35}. 

Gauge invariance of \eqref{B.35a} requires $\Lambda_{\a}$ to also have pseudo-mass $M$,
\begin{align}
0&=\big(\mathbb{Q}- M^2\big)\Lambda_{\a}~. \label{B.37}
\end{align}
Next we decompose the gauge parameter into irreducible parts. Performing this procedure in accordance with the discussion in section \ref{decomp}, we find that \eqref{B.36} takes the form
\begin{align}
\delta H_{\a\ad}=\bar{\mathcal{D}}_{\ad}\zeta_{\a}-\mathcal{D}_{\a}\bar{\z}_{\ad}+\mathcal{D}_{\a\ad}\xi + \big[\mathcal{D}_{\a},\bar{\mathcal{D}}_{\ad}\big]\z+\mathcal{D}_{\a\ad}\big(\s+\bar{\s}\big)~. \label{B.38}
\end{align}
In \eqref{B.38}, the gauge parameter $\z_{\a}$ is LTAL
\begin{align}
0=\big(\bar{\mathcal{D}}^2-4\mu\big)\z_{\a}~,\qquad 0=\mathcal{D}^{\a}\z_{\a}~~ \Leftrightarrow ~~ 0=\big(\mathcal{D}^2-6\mub\big)\z_{\a}~. \label{B.39}
\end{align}
The real parameters $\xi=\bar{\xi}$ and $\z=\bar{\zeta}$ are linear
\begin{align}
0=\big(\bar{\mathcal{D}}^2-4\mu\big)\z=\big(\mathcal{D}^2-4\mub\big)\z~,\qquad  0=\big(\bar{\mathcal{D}}^2-4\mu\big)\xi=\big(\mathcal{D}^2-4\mub\big)\xi~, \label{B.40}
\end{align}
whilst $\s$ is complex chiral, $\bar{\mathcal{D}}_{\ad}\s=0$.

Before solving the equation $0=\mathcal{D}^{\b}\delta H_{\b\ad}$, we first analyse the higher-order equations $0=\mathcal{D}^{\b\bd}\delta H_{\b\bd}$ and $0=\big[\mathcal{D}^{\b},\bar{\mathcal{D}}^{\bd}\big]\delta H_{\b\bd}$. Respectively,  they yield the following equations\footnote{The (non-unitary) masses $M^2=2\mu\mub$ and $M^2=0$ should be considered separately. The result of this analysis is that there is no gauge-symmetry present.}
\begin{subequations}\label{B.41}
\begin{align}
0&=\big(M^2-\frac{1}{4}\mu\mathcal{D}^2-\frac{1}{4}\mub\bar{\mathcal{D}}^2\big)(\s+\bar{\s})+\big(M^2-2\mu\mub\big)\xi~,\label{B.41a}\\
0&=\big(M^2+\frac{3}{4}\mu\mathcal{D}^2+\frac{3}{4}\mub\bar{\mathcal{D}}^2\big)(\s-\bar{\s})-6\text{i}\big(M^2-2\mu\mub\big)\zeta~.\label{B.41b}
\end{align}
\end{subequations}
One may show that both equations in \eqref{B.41} are consistent with \eqref{B.40} only if 
\begin{align}
\zeta=0~,\qquad \xi=0~,
\end{align}
which holds for all values of $M^2$. In addition, we must also have 
\begin{align}
\sigma = \bar{\s}=0~,
\end{align}
unless $M^2=M^2_{(2)}$, in which case $\s$ must satisfy the equation 
\begin{align}
0=-\frac{1}{4}(\mathcal{D}^2-4\mub )\s + 2\mub \bar{\s}~. \label{B.43}
\end{align}
Finally, using the above information to solve the equation $0=\mathcal{D}^{\b}\delta H_{\b\ad}$, one arrives at the reality conditions (valid for all $M^2$)
\begin{align}
\mathcal{D}_{\ad}{}^{\b}\z_{\b}=2\text{i}\mub \bar{\z}_{\ad}~,\qquad \mathcal{D}_{\a}{}^{\bd}\bar{\z}_{\bd}=-2\text{i}\mu \z_{\a}~, \label{B.44}
\end{align}
which imply the mass-shell condition
\begin{align}
0=\big(M^2-5\mu\mub\big)\z_{\a}~.
\end{align}
We see that unless $H_{\a\ad}$ has pseudo-mass  $M^2=M^2_{(1)}$, then
\begin{align}
\zeta_{\a}=0~.
\end{align}

In conclusion, the on-shell conditions \eqref{B.35} are only compatible with a gauge symmetry when the pseudo-mass of $H_{\a\ad}$ (and the corresponding gauge parameters) takes one of the partially-massless values \eqref{B.355a} or \eqref{B.355b}. For super-depth $t=1$, the relevant gauge symmetry is
\begin{align}
\delta H_{\a\ad}=\bar{\mathcal{D}}_{\ad}\zeta_{\a}-\mathcal{D}_{\a}\bar{\z}_{\ad}~,
\end{align}
where $\z_{\a}$ satisfies \eqref{B.39} and \eqref{B.44}. On the otherhand, for super-depth $t=2$ it is
\begin{align}
\delta H_{\a\ad}= \mathcal{D}_{\a\ad}(\s+\bar{\s})~,
\end{align}
where $\s$ is chiral and satisfies \eqref{B.43}.

By conducting a similar analysis on the real supermultiplet $H_{\a(2)\ad(2)}$, one arrives at the gauge transformations presented in table \ref{table 3}. In principle this procedure may be carried out for any on-shell superfield $\phi_{\a(m)\ad(n)}$ with fixed $m$ and $n$, 
but in practice this procedure is quite time consuming. In section \ref{section 3.2} we use the above analysis to motivate an ansatz for the case arbitrary half-integer superspin.

\begin{footnotesize}
	
\end{footnotesize}

\begin{thebibliography}{66}
		

\bibitem{SalamS} 
  A.~Salam and J.~A.~Strathdee,
  ``On superfields and Fermi-Bose symmetry,''
  Phys.\ Rev.\ D {\bf 11}, 1521 (1975).

\bibitem{Sokatchev} 
  E.~Sokatchev,
  ``Projection operators and supplementary conditions for superfields 
  with an arbitrary spin,''
  Nucl.\ Phys.\ B {\bf 99}, 96 (1975).
  
\bibitem{SG} 
  W.~Siegel and S.~J.~Gates Jr.,
 ``Superprojectors,''
  Nucl.\ Phys.\ B {\bf 189}, 295 (1981).
  
\bibitem{Sokatchev81}
E.~Sokatchev,
``Irreducibility conditions for extended superfields,''
Phys. Lett. B \textbf{104}, 38-40 (1981).
 
  
 \bibitem{RS}
V.~Rittenberg and E.~Sokatchev,
``Decomposition of extended superfields into irreducible representations of supersymmetry,''
Nucl. Phys. B \textbf{193}, 477-501 (1981).


\bibitem{GGRS}
 S.~J.~Gates Jr., M.~T.~Grisaru, M.~Ro\v{c}ek and W.~Siegel,
{\it Superspace, or One Thousand and One Lessons in Supersymmetry},
Benjamin/Cummings (Reading, MA), 1983,
arXiv:hep-th/0108200.
     


\bibitem{BHHK}
E.~I.~Buchbinder, D.~Hutchings, J.~Hutomo and S.~M.~Kuzenko,
``Linearised actions for $ \mathcal{N} $-extended (higher-spin) superconformal gravity,''
JHEP \textbf{08} (2019) 077
[arXiv:1905.12476 [hep-th]].
 

\bibitem{OS1} 
  V.~Ogievetsky and E.~Sokatchev,
  ``On vector superfield generated by supercurrent,''
  Nucl.\ Phys.\ B {\bf 124}, 309 (1977).

\bibitem{OS2} 
  V.~I.~Ogievetsky and E.~Sokatchev,
  ``Superfield equations of motion,''
  J.\ Phys.\ A {\bf 10}, 2021 (1977).
  
\bibitem{GS}
S.~J.~Gates Jr. and W.~Siegel,
``(3/2, 1) superfield of O(2) supergravity,''
Nucl.\ Phys.\ B {\bf 164},  484 (1980).  


\bibitem{GKP}
  S.~J.~Gates Jr., S.~M.~Kuzenko and J.~Phillips,
  ``The off-shell (3/2,2) supermultiplets revisited,''
  Phys.\ Lett.\  B {\bf 576}, 97 (2003)  [arXiv:hep-th/0306288].
  
\bibitem{BKLP}
E.~I.~Buchbinder, S.~M.~Kuzenko, J.~La Fontaine and M.~Ponds,
``Spin projection operators and higher-spin Cotton tensors in three dimensions,''
Phys. Lett. B \textbf{790}, 389-395 (2019)
[arXiv:1812.05331 [hep-th]].

\bibitem{BF} 
R.~E.~Behrends and C.~Fronsdal,
``Fermi decay of higher spin particles,'' Phys.\ Rev.\  {\bf 106},  345 (1957).

\bibitem{Fronsdal}
C. Fronsdal, ``On the theory of higher spin fields,'' 
Nuovo Cim. {\bf  9}, 416 (1958).


\bibitem{IS}
E.~A.~Ivanov and A.~S.~Sorin,
``Superfield formulation of OSp(1,4) supersymmetry,''
J.\ Phys.\ A  {\bf 13} (1980) 1159.


		
\bibitem{KS94}
S.~M.~Kuzenko and A. G. Sibiryakov,
``Free massless higher-superspin superfields on the	anti-de Sitter superspace,''
Phys. Atom. Nucl. \textbf{57}, 1257-1267 (1994)
	[arXiv:1112.4612 [hep-th]].


\bibitem{KPS}
  S.~M.~Kuzenko, A.~G.~Sibiryakov and V.~V.~Postnikov,
  ``Massless gauge superfields of higher half integer superspins,''
  JETP Lett.\  {\bf 57},  534 (1993)  [Pisma Zh.\ Eksp.\ Teor.\ Fiz.\  {\bf 57},  521 (1993)].

\bibitem{KS}
  S.~M.~Kuzenko and A.~G.~Sibiryakov,
  ``Massless gauge superfields of higher integer superspins,''
  JETP Lett.\  {\bf 57},  539 (1993)   [Pisma Zh.\ Eksp.\ Teor.\ Fiz.\  {\bf 57},  526 (1993)].

\bibitem{Sibiryakov} A. G. Sibiryakov,
``Superfield models for the massless higher-superspin multiplets,'' Ph.D. thesis, Tomsk State University (1996). 


\bibitem{KR}
S.~M.~Kuzenko and E.~S.~N.~Raptakis,
``Symmetries of supergravity backgrounds and supersymmetric field theory,''
JHEP {\bf 2004}, 133 (2020)
  [arXiv:1912.08552 [hep-th]].

\bibitem{FZ}
 S.~Ferrara and B.~Zumino,
``Transformation properties of the supercurrent,''
Nucl.\ Phys.\  B {\bf 87}, 207 (1975).


\bibitem{HST81}
P.~S.~Howe, K.~S.~Stelle and P.~K.~Townsend,
``Supercurrents,''  Nucl.\ Phys.\  B {\bf 192}, 332 (1981).


\bibitem{KMT} 
S.~M.~Kuzenko, R.~Manvelyan and S.~Theisen,
``Off-shell superconformal higher spin multiplets in four dimensions,''
JHEP {\bf 1707}, 034 (2017)
[arXiv:1701.00682 [hep-th]].




\bibitem{BHK} 
E.~I.~Buchbinder, J.~Hutomo and S.~M.~Kuzenko,
``Higher spin supercurrents in anti-de Sitter space,''
JHEP {\bf 1809}, 027 (2018)
[arXiv:1805.08055 [hep-th]].  


\bibitem{KT}
S.~M.~Kuzenko and S.~Theisen,
``Correlation functions of conserved currents in N = 2 superconformal
theory,''  Class.\ Quant.\ Grav.\  {\bf 17}, 665 (2000)  [hep-th/9907107].


\bibitem{Sohnius79}
  M.~F.~Sohnius,
  ``The multiplet of currents for N=2 extended supersymmetry,''
  Phys.\ Lett.\  B {\bf 81}, 8 (1979).


\bibitem{FWZ} 
S.~Ferrara, J.~Wess and B.~Zumino,
``Supergauge multiplets and superfields,''
  Phys.\ Lett.\ B {\bf 51}, 239 (1974).
	
\bibitem{BKS}
I.~L.~Buchbinder, S.~M.~Kuzenko and A.~G.~Sibiryakov,
``Quantization of higher spin superfields in the anti-de Sitter superspace,''
Phys. Lett. B \textbf{352}, 29-36 (1995)
[arXiv:hep-th/9502148 [hep-th]].

\bibitem{KP20}
S.~M.~Kuzenko and M.~Ponds,
``Spin projection operators in (A)dS and partial masslessness,''
Phys. Lett. B \textbf{800}, 135128 (2020)
[arXiv:1910.10440 [hep-th]].


\bibitem{DW2} 
  S.~Deser and A.~Waldron,
  ``Partial masslessness of higher spins in (A)dS,''
  Nucl.\ Phys.\ B {\bf 607}, 577 (2001)
  [hep-th/0103198].  

\bibitem{DeserN1} 
  S.~Deser and R.~I.~Nepomechie,
  ``Anomalous propagation of gauge fields in conformally flat spaces,''
  Phys.\ Lett.\  {\bf 132B}, 321 (1983).
  
\bibitem{DeserN2}
  S.~Deser and R.~I.~Nepomechie,
  ``Gauge invariance versus masslessness in de Sitter space,''
  Annals Phys.\  {\bf 154} (1984) 396.
  
\bibitem{Higuchi1} 
  A.~Higuchi,
  ``Forbidden mass range for spin-2 field theory in de Sitter space-time,''
  Nucl.\ Phys.\ B {\bf 282}, 397 (1987).
  
  \bibitem{Higuchi2} 
  A.~Higuchi,
  ``Symmetric tensor spherical harmonics on the $N$ sphere and their application to the de Sitter group SO($N$,1),''
  J.\ Math.\ Phys.\  {\bf 28}, 1553 (1987)
  
  
  \bibitem{Higuchi3} 
  A.~Higuchi,
  ``Massive symmetric tensor field in space-times with a positive cosmological constant,''
  Nucl.\ Phys.\ B {\bf 325}, 745 (1989).
  
  
  \bibitem{DeserW1} 
  S.~Deser and A.~Waldron,
  ``Gauge invariances and phases of massive higher spins in (A)dS,''
  Phys.\ Rev.\ Lett.\  {\bf 87}, 031601 (2001)
  [hep-th/0102166].
  
  
 \bibitem{DeserW3} 
  S.~Deser and A.~Waldron,
  ``Stability of massive cosmological gravitons,''
  Phys.\ Lett.\ B {\bf 508}, 347 (2001)
  [hep-th/0103255]. 

\bibitem{DeserW4} 
  S.~Deser and A.~Waldron,
  ``Null propagation of partially massless higher spins in (A)dS and cosmological constant speculations,''
  Phys.\ Lett.\ B {\bf 513}, 137 (2001)
  [hep-th/0105181].
  
 \bibitem{Zinoviev} 
  Y.~M.~Zinoviev,
  ``On massive high spin particles in AdS,''
  hep-th/0108192. 

\bibitem{Metsaev2}
L.~Brink, R.~R.~Metsaev and M.~A.~Vasiliev,
``How massless are massless fields in AdS(d),''
Nucl. Phys. B \textbf{586}, 183 (2000),
[arXiv:hep-th/0005136 [hep-th]].

 \bibitem{SV} 
  E.~D.~Skvortsov and M.~A.~Vasiliev,
  ``Geometric formulation for partially massless fields,''
  Nucl.\ Phys.\ B {\bf 756}, 117 (2006)
  [hep-th/0601095].  


  \bibitem{Metsaev}
  R.~R.~Metsaev,
 ``Gauge invariant formulation of massive totally symmetric fermionic fields in (A)dS space,''
  Phys.\ Lett.\ B {\bf 643} (2006) 205
  [hep-th/0609029].
  

\bibitem{BG}
X.~Bekaert and M.~Grigoriev,
``Higher order singletons, partially massless fields and their boundary values in the ambient approach,''
Nucl. Phys. B \textbf{876}, 667-714 (2013)
[arXiv:1305.0162 [hep-th]].


 
  
 \bibitem{G-SHR}
S.~Garcia-Saenz, K.~Hinterbichler and R.~A.~Rosen,
``Supersymmetric partially massless fields and non-unitary 
superconformal representations,''
JHEP \textbf{11}, 166 (2018)
[arXiv:1810.01881 [hep-th]].
 
 \bibitem{BKSZ}
I.~L.~Buchbinder, M.~V.~Khabarov, T.~V.~Snegirev and Y.~M.~Zinoviev,
``Lagrangian description of the partially massless higher spin N = 1 supermultiplets in AdS$_{4}$ space,''
JHEP \textbf{08}, 116 (2019)
[arXiv:1904.01959 [hep-th]].


\bibitem{Dirac:1935zz}
P.~A.~M.~Dirac,
``The electron wave equation in De-Sitter space,''
Annals Math. \textbf{36}, 657-669 (1935).

\bibitem{Dirac:1963ta}
P.~A.~M.~Dirac,
``A remarkable representation of the 3 + 2 de Sitter group,''
J. Math. Phys. \textbf{4}, 901-909 (1963).


\bibitem{Fronsdal:1965zzb}
C.~Fronsdal,
``Elementary particles in a  curved space,''
Rev. Mod. Phys. \textbf{37}, 221-224  (1965).


\bibitem{Fronsdal:1974ew}
C.~Fronsdal,
``Elementary particles in a curved space. ii,''
Phys. Rev. D \textbf{10}, 589-598 (1974).


\bibitem{Fronsdal:1975eq}
C.~Fronsdal and R.~B.~Haugen,
``Elementary particles in a curved space. 3,''
Phys. Rev. D \textbf{12}, 3810-3818  (1975).


\bibitem{Fronsdal:1975ac}
C.~Fronsdal,
``Elementary particles in a curved space. 4. Massless particles,''
Phys. Rev. D \textbf{12}, 3819 (1975).


\bibitem{Evans} N.~T.~Evans, ``Discrete series for the universal covering group of the 3 + 2 de Sitter group,"
J. Math. Phys. {\bf 8},  170 (1967). 
			
			
\bibitem{Angelopoulos}  E. Angelopoulos, ``$\overline{\sSO}_0 (3, 2)$: Linear and unitary irreducible representations,'' in 
{\it Quantum Theory, Groups, Fields and Particles}, A. O. Barut (Ed.),
D. Reidel Publishing, 1983, pp 101--148.

\bibitem{AFFS}
E.~Angelopoulos, M.~Flato, C.~Fronsdal and D.~Sternheimer,
``Massless particles, conformal group and de Sitter universe,''
Phys. Rev. D \textbf{23}, 1278 (1981).


\bibitem{Heidenreich:1982rz}
W.~Heidenreich,
``All linear unitary irreducible representations of de Sitter supersymmetry with positive energy," Phys. Lett. B \textbf{110}, 461-464 (1982).

\bibitem{Nicolai:1984hb}
H.~Nicolai,
"Representations of supersymmetry in anti-de Sitter space,"
in: {\it Supersymmetry and Supergravity '84}, Proceedings of the Trieste Spring School, eds. B. de Wit, P. Fayet, P. van Nieuwehuizen (Worlds Scientific, 1984).

\bibitem{deWit:1999ui}
B.~de Wit and I.~Herger,
``Anti-de Sitter supersymmetry,''
Lect. Notes Phys. \textbf{541}, 79--100  (2000)
[arXiv:hep-th/9908005 [hep-th]].

\bibitem{FF78}
M.~Flato and C.~Fronsdal,
``One massless particle equals two Dirac singletons: 
Elementary particles in a curved space. 6,''
Lett. Math. Phys. \textbf{2}, 421-426 (1978).


\bibitem{Flato:1980zk}
M.~Flato and C.~Fronsdal,
``On Dis and Racs,''
Phys. Lett. B \textbf{97}, 236-240 (1980).

\bibitem{Barut:1970kp}
A.~O.~Barut and A.~Boehm,
``Reduction of a class of o(4,2) representations with respect to so(4,1) and so(3,2),''
J. Math. Phys. \textbf{11}, 2938-2945 (1970).

\bibitem{Fronsdal:1985pc}
C.~Fronsdal,
``3+2 de Sitter superfields,''
Math. Phys. Stud. \textbf{8}, 67-122 (1986).

\bibitem{Fronsdal:1981gq}
C.~Fronsdal,
``The Dirac supermultiplet,''
Phys. Rev. D \textbf{26}, 1988 (1982).

\bibitem{BreitenF}
P.~Breitenlohner and D.~Z.~Freedman,
``Stability in gauged extended supergravity,''
Annals Phys. \textbf{144}, 249 (1982).

\bibitem{Keck}
  B.~W.~Keck,
 ``An alternative class of supersymmetries,''
J.\ Phys.\ A  {\bf 8}, 1819 (1975).

\bibitem{Zumino77}
B.~Zumino, ``Nonlinear realization of supersymmetry in de Sitter space,''
Nucl.\ Phys.\  B {\bf 127}, 189 (1977).

\bibitem{BK} 
I.~L. Buchbinder and S.~M. Kuzenko, {\it Ideas and Methods of Supersymmetry and
Supergravity, Or a Walk Through Superspace},
 IOP, Bristol, 1995 (Revised Edition 1998).

\bibitem{ButterK} 
  D.~Butter and S.~M.~Kuzenko,
  ``A dual formulation of supergravity-matter theories,''
  Nucl.\ Phys.\ B {\bf 854}, 1 (2012)
  [arXiv:1106.3038 [hep-th]].

\bibitem{KP19}
S.~M.~Kuzenko and M.~Ponds,
``Conformal geometry and (super)conformal higher-spin gauge theories,''
JHEP {\bf 1905},  113 (2019) 
[arXiv:1902.08010 [hep-th]].


\bibitem{Siegel78}
W.~Siegel,
``Solution to constraints in Wess-Zumino supergravity formalism,''
Nucl.\ Phys.\  B {\bf 142}, 301 (1978). 

\bibitem{KPR}
S.~M.~Kuzenko, M.~Ponds and E.~S.~N.~Raptakis,
``New locally (super)conformal gauge models in Bach-flat backgrounds,''
JHEP \textbf{08}, 068 (2020)
[arXiv:2005.08657 [hep-th]].


\bibitem{FT} 
E.~S.~Fradkin and A.~A.~Tseytlin,  ``Conformal supergravity,''
Phys.\ Rept.\  {\bf 119}, 233 (1985).

\bibitem{Vasiliev} 
  M.~A.~Vasiliev,
  ``Free massless fields of arbitrary spin in the de Sitter space and initial data 
  for a higher spin superalgebra,''
  Fortsch.\ Phys.\  {\bf 35}, 741 (1987).
  
  \bibitem{KPR2}
S.~M.~Kuzenko, M.~Ponds and E.~S.~N.~Raptakis,
``Generalised superconformal higher-spin multiplets,''
[arXiv:2011.11300 [hep-th]].


\bibitem{BFDG}
C.~J.~C.~Burges, D.~Z.~Freedman, S.~Davis and G.~W.~Gibbons,
``Supersymmetry in anti-de Sitter space,''
Annals Phys. \textbf{167}, 285 (1986).

\bibitem{DF}
D.~W.~D\"usedau and D.~Z.~Freedman,
``Renormalization in anti-de Sitter supersymmetry,''
Phys. Rev. D \textbf{33}, 395 (1986).

\bibitem{BG1}
S.~Bellucci and J.~Gonzalez,
``One-loop order renormalization of the massive \{Wess-Zumino\} model in anti-de Sitter space,''
Phys. Rev. D \textbf{33}, 2319 (1986).

\bibitem{BG2}
S.~Bellucci and J.~Gonzalez,
``Superfield formulation of the Wess-Zumino model in anti-de Sitter space,''
Class. Quant. Grav. \textbf{6}, 505 (1989).

\bibitem{Siegel-tensor}
W.~Siegel,
``Gauge spinor superfield as a scalar multiplet,''
Phys.\ Lett.\ B {\bf 85}, 333 (1979).

\bibitem{VanProeyen}
A.~Van Proeyen,
``Massive vector multiplets in supergravity,''
Nucl. Phys. B \textbf{162},  376 (1980). 


\bibitem{AB}
R.~Altendorfer and J.~Bagger,
``Dual anti-de Sitter superalgebras from partial supersymmetry breaking,''
Phys. Rev. D \textbf{61}, 104004 (2000)
[arXiv:hep-th/9908084 [hep-th]].		


\bibitem{Zinoviev07}
Y.~M.~Zinoviev,
``Massive supermultiplets with spin 3/2,''
JHEP \textbf{05}, 092 (2007)
[arXiv:hep-th/0703118 [hep-th]].

\bibitem{Butter:2011ym}
D.~Butter and S.~M.~Kuzenko,
``N=2 AdS supergravity and supercurrents,''
JHEP \textbf{07}, 081 (2011)
[arXiv:1104.2153 [hep-th]].

\bibitem{BGKP}
I.~L.~Buchbinder, S.~J. Gates Jr., S.~M.~Kuzenko and J.~Phillips,
``Massive 4D, N=1 superspin 1 \& 3/2 multiplets and dualities,''
JHEP \textbf{02}, 056 (2005)
[arXiv:hep-th/0501199 [hep-th]].

\bibitem{Tseytlin5}
  A.~A.~Tseytlin,
  ``Effective action in de Sitter space and conformal supergravity,''
   Yad.\ Fiz.\  {\bf 39}, 1606 (1984)
  [Sov.\ J.\ Nucl.\ Phys.\  {\bf 39}, no. 6, 1018 (1984)].

\bibitem{Tseytlin6}
  E.~S.~Fradkin and A.~A.~Tseytlin,
  ``Instanton zero modes and beta functions in supergravities. 2. Conformal supergravity,''
  Phys.\ Lett.\  {\bf 134B} (1984) 307.
  
\bibitem{Tseytlin13} 
  A.~A.~Tseytlin,
  ``On partition function and Weyl anomaly of conformal higher spin fields,''
  Nucl.\ Phys.\ B {\bf 877}, 598 (2013)
  [arXiv:1309.0785 [hep-th]].  
   
   
   
\bibitem{Karapet1}
  E.~Joung and K.~Mkrtchyan,
  ``A note on higher-derivative actions for free higher-spin fields,''
  JHEP {\bf 1211} (2012) 153
  [arXiv:1209.4864 [hep-th]].
   
   
\bibitem{Karapet2}
  E.~Joung and K.~Mkrtchyan,
  ``Weyl action of two-column mixed-symmetry field and its factorization around (A)dS space,''
  JHEP {\bf 1606} (2016) 135
  [arXiv:1604.05330 [hep-th]].
   
\bibitem{Metsaev2014} 
  R.~R.~Metsaev,
  ``Arbitrary spin conformal fields in (A)dS,''
  Nucl.\ Phys.\ B {\bf 885}, 734 (2014)
  [arXiv:1404.3712 [hep-th]]. 
 
\bibitem{NT} 
  T.~Nutma and M.~Taronna,
  ``On conformal higher spin wave operators,''
  JHEP {\bf 1406}, 066 (2014)
  [arXiv:1404.7452 [hep-th]].  

\bibitem{GH} 
  M.~Grigoriev and A.~Hancharuk,
  ``On the structure of the conformal higher-spin wave operators,''
  JHEP {\bf 1812}, 033 (2018)
  [arXiv:1808.04320 [hep-th]].

	
\end{thebibliography}
\end{document}